%% file: FTtoLBM.tex
\documentclass[a4paper,12pt,english]{article}
\usepackage[dvipdfm]{color,graphicx}
\input revmacro

\usepackage{babel}

\newcommand {\beq}{\begin{equation}}
\newcommand {\eeq}{\end{equation}}
\newcommand {\bea}{\begin{eqnarray}}
\newcommand {\eea}{\end{eqnarray}}
\newcommand {\nn}{\nonumber \\}

\newcommand {\pl}{\partial}

\newcommand {\al}{\alpha}
\newcommand {\be}{\beta}

\newcommand {\Ga}{\Gamma}

\newcommand {\ka}{\kappa}
\newcommand {\la}{\lambda}
\newcommand {\La}{\Lambda}
\newcommand {\si}{\sigma}

\newcommand {\sh}{\theta}   

\newcommand {\om}{\omega}
\newcommand {\Om}{\Omega}

\newcommand {\ep}{\epsilon}
\newcommand {\vep}{\varepsilon}

\newcommand {\del}  {\delta}
\newcommand {\Del}  {\Delta}

\newcommand {\Tcal}{{\cal T}}
\newcommand {\Dcal}{{\cal D}}

\newcommand {\Wcal}{{\cal W}}

\newcommand {\Dtil}{{\tilde D}}

\newcommand {\ftil}{{\tilde f}}
\newcommand {\Itil}{{\tilde I}}

\newcommand {\ltil} {{\tilde l}}
\newcommand {\ttil} {{\tilde t}}

\newcommand {\tautil} {{\tilde \tau}}

\newcommand {\rhotil} {{\tilde \rho}}

\newcommand {\Nbar}  {{\bar N}}

\newcommand {\Vbar}  {{\bar V}}

\newcommand {\lbar}{\bar{l}}

\newcommand {\rdot}{\dot{r}}

\newcommand {\wdot}{\dot{w}}
\newcommand {\xdot}{\dot{x}}
\newcommand {\ydot}{\dot{y}}

\newcommand {\xddot}{\ddot{x}}

\newcommand {\bu}{{\bf u}}
\newcommand {\bv}{{\bf v}}
\newcommand {\bx}{{\bf x}}

\newcommand {\bna}{{\bf \nabla}}

\newcommand {\etap}{{\eta'}}

\newcommand {\intthx} {{\int d^3x}}

\newcommand {\intx} {{\int dx}}

\newcommand {\ra} {\rightarrow}

\newcommand {\ul}   {\underline}
\newcommand {\pr}   {{\quad .}}
\newcommand {\com}  {{\quad ,}}
\newcommand {\q}    {\quad}
\newcommand {\qq}   {\quad\quad}
\newcommand {\qqq}   {\quad\quad\quad}
\newcommand {\qqqq}   {\quad\quad\quad\quad}
\newcommand {\qqqqq}   {\quad\quad\quad\quad\quad}

\newcommand {\nl}    {\newline}

\newcommand {\NP}   {Nucl.Phys.}
\newcommand {\PL}   {Phys.Lett.}
\newcommand {\PR}   {Phys.Rev.}

\newcommand {\PTP}  {Prog.Theor.Phys.}

\newcommand {\uinf} {{u^\infty}}
\newcommand {\finf} {{f^\infty}}
\newcommand {\tn} {{t_n}}
\newcommand {\un} {{u_n}}
\newcommand {\xn} {{x_n}}
\newcommand {\fn} {{f_n}}
\newcommand {\phin} {{\phi_n}}
\newcommand {\rhon} {{\rho_n}}
\newcommand {\alinv} {{\frac{1}{\alpha}}}
\newcommand {\sqral} {{\sqrt{\alpha}}}
\newcommand {\tauzero} {{\tau_0}}
\newcommand {\kB} {{k_B}}

\begin{document}

\title{
Velocity-Field Theory, Boltzmann's Transport Equation, Geometry  
and Emergent Time 
      }
\author{Shoichi Ichinose
\footnote{
E-mail: ichinose@u-shizuoka-ken.ac.jp
             }
           }

\maketitle
\begin{center}\emph{
Laboratory of Physics, School of Food and Nutritional Sciences, 
University of Shizuoka, 
Yada 52-1, Shizuoka 422-8526, Japan
                           }
\end{center}

\begin{abstract}
Boltzmann equation describes the time development of 
the velocity distribution in the continuum fluid matter. 
We formulate the equation using the field theory where 
the {\it velocity-field} plays the central role. The properties of the  
fluid matter (fluid particles)  
appear as the density and the viscosity.  
{\it Statistical fluctuation} is examined, and is clearly discriminated from the quantum effect. 
The time variable is {\it emergently} introduced through the computational process step. 
Besides the ordinary potential, the general velocity potential is introduced. 
The collision term, for  
the Higgs-type velocity potential, is explicitly obtained and 
the (statistical) fluctuation is closely explained. 
The system is generally {\it non-equilibrium}. 
The present field theory model does {\it not} conserve energy and is an open-system model. 
One dimensional Navier-Stokes equation, i.e., Burgers equation,  appears. 
In the latter part of the text, we present a way to directly define the 
distribution function by use of the geometry, appearing in the energy expression, 
and Feynman's path-integral. 
\end{abstract}

Key Words\ :\  Boltzmann equation;\ 
velocity field theory;\ 
statistical fluctuation;\ 
computational step number;\ 
open system;\ 
geometry.

\section{Introduction\label{intro}}
Boltzmann equation was introduced to explain the second law of 
the thermodynamics in the dynamical way, in 1872, by Boltzmann. 
We considers the (visco-elastic) fluid matter and examine the dynamical behavior using the velocity-field theory. 
The scale size we consider is far bigger than the atomic scale ($\sim 10^{-10}$m) and 
is smaller than or nearly equal to the optical microscope scale ($\sim 10^{-6}$m). 
The equation describes the 
temporal development of 
the distribution function $f(t,\bx,\bv)$ which shows the probability 
of a fluid-molecule (particle) having the velocity $\bv$ at the space $\bx$ and time $t$. 

 We reformulate the Boltzmann equation using the field theory of the velocity 
field $\bu(\bx,~'t')$. Basically it is based on the {\it minimal energy principle}. 
We do {\it not} introduce time $t$. Instead of $t$, we use 
the {\it computational step number} $n$.  The system 
we consider consists of the huge number of fluid-particles (molecules) and the 
physical quantities, such as energy and entropy, are the {\it statistically-averaged} ones. 
It is not obtained by the deterministic way like the classical (Newton) mechanics. 
We introduce the {\it statistical ensemble} by 
using the well-established field-theory method, 
the {\it background-field method}\cite{DeW67, tHooft73}. 
Renormalization phenomenon  occurs not from the quantum effect but from 
the statistical fluctuation due to 
the inevitable uncertainty caused by 1) the step-wise (discrete-time) formulation and 
2) the continuum formulation for the space (of the real matter world). 

After the development of the string and D-brane theories\cite{StringText1, StringText2}, 
one general relation, between the 4-dimensional(4D) conformal theories and the 5D gravitational 
theories, was proposed. The 5D gravitational theories are asymptotically AdS$_5$\cite{Malda9711,GKP9802,Witten9802}. 
The proposal claims the quantum behavior of the 4D theories is obtainable 
by the classical analysis of the 5D gravitational ones. The development along the extra axis can be 
regarded as the renormalization flow. This approach (called AdS/CFT) has been providing 
non-perturbative studies in several branches: quark-gluon plasma physics, heavy-ion collisions, 
non-equilibrium statistical mechanics, superconductivity, superfluidity\cite{Natsu2015, AE2015}. 
Especially, as the most relevant 
to the present work, the connection with the hydrodynamics is important\cite{Natsu08}. 
When a black hole is given a perturbation, the effect decays as the relaxation phenomenon. The transport coefficients, such as viscosities, speed of sound, 
thermal conductivity, are important physical quantities.

The dissipative system we consider is characterized by the dissipation of energy (heat). 
Even for the particle classical (Newton) mechanics, the notion of energy is somewhat obscure 
when the dissipation occurs. We consider the movement of a particle 
under the influence of the friction force, $F_{\mbox{friction}}$. The emergent energy, 
$E(x(t), \xdot(t))$, during the period 
[t$_1$, t$_2$] can {\it not} be written as the following form. 
\bea
\int_{x_1}^{x_2}F_{\mbox{friction}}~dx=\left[ E(x(t), \xdot(t)) \right]_{t_1}^{t_2}= 
E|_{t_2}-E|_{t_1}\ ,\ 
x_1=x(t_1)\ ,\ x_2=x(t_2)
\ ,
\label{intro1}
\eea 
where $x(t)$ is the orbit (path) of the particle. It depends on the path (or orbit) itself. It cannot be written as the form of difference between some quantity ($E(t)$) at 
time t$_1$ and t$_2$. In this situation, we realize the time itself should be re-considered 
when the dissipation occurs. Owing to Einstein's idea of "space-time democracy", we have 
stuck to the standpoint that space and time should be treated on the equal footing. We 
present here the {\it step-wise} approach to the time-development. 

 We do {\it not} use time variable. Instead we use the computational-process step number $n$. 
Hence the {\it increasing} of the number $n$ is identified as the {\it time development}. 
The connection between step $n$ and step $n-1$ is determined by 
the {\it minimal energy principle}. In this sense, time is "emergent" from 
the minimal energy principle. The direction of flow (arrow of time) is built in from the beginning. 
\footnote{
For a recent review on the nature of time, see ref.\cite{Gibbons11}.  
}
The usefulness of the step number approach (discrete Morse flow method) is shown  
in the text. The step-wise treatment makes the 'time' direction manipulation 
more skillful and fruitful. 

In the latter part of this paper, an approach to the statistically-
averaging procedure 
, based on the {\it geometry} of the mechanical dynamics, is presented. 

The content is described as follows. The step-wise dynamical equation is presented 
in Sec.2. We start with the n-th step energy functional. By regarding n steps as 
the time t$_n$, we derive Burgers equation (1 dim Navier-Stokes equation). 
In Sec.3, the orbit (path) 
of the fluid particle is explained in this step-wise formalism. The total energy and 
the energy rate are also explained. The statistical fluctuation is closely explained 
in Sec.4. Especially the difference from the quantum effect is stressed. Using the 
path-integral, we take into account the fluctuation effect. Owing to the present 
velocity-field formalism, we can obtain Boltzmann's equation, as described in Sec.5, 
up to the collision term. This step-wise approach is applied to the mechanical system 
in Sec.6. We take a simple dissipative model:\ the harmonic oscillator with friction. 
The trajectory is solved in the step-wise way. We find the total energy changes 
as the step proceeds. From the n-step energy expression we can extract 
the geometrical structure (the metric) of the trajectory.  
The metric is used, in Sec.7, to define the {\it statistical ensemble} of the system of 
N viscous particles in one space dimension. We propose some models using the geometrically-basic quantities: 
the length and the area. Conclusion is given in Sec.\ref{conc}. Some appendices are 
provided to supplement the text. App.A treats (1+3) dimensional field theory in this 
step-wise formalism. App.B is the calculation of the statistical fluctuation effect 
and it supplements Sec.\ref{fluct}. 
A few simulation results of the frictional harmonic oscillator (Sec.\ref{qm}) are shown 
in App.C. 
An additional mechanical model (Spring-Block model) 
is described with the calculation result in App.D.

\vspace{5mm}
\begin{figure}
\caption{
The energy functional $I_n[u(x)]$, (\ref{DTheat1}), of the velocity-field $u(x)$. 
        }
\begin{center}
\includegraphics[height=8cm]{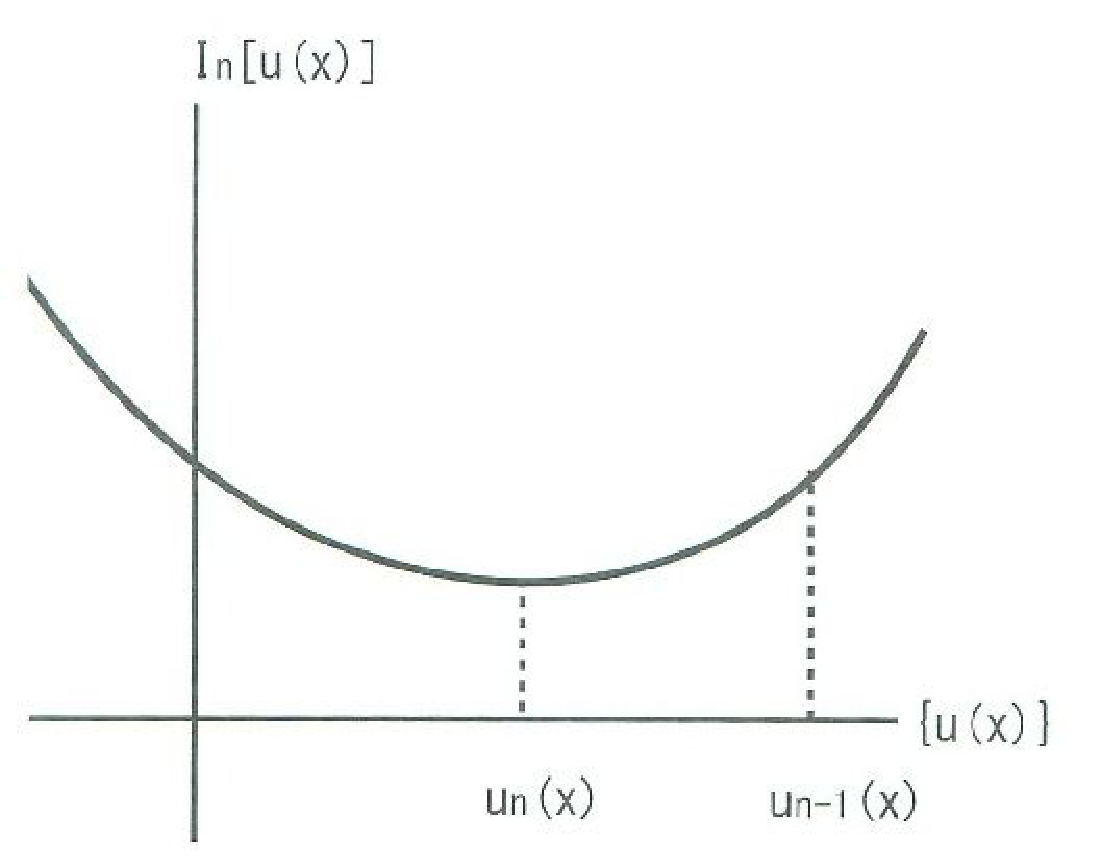}
\end{center}
\label{EneFunc}
\end{figure}
\section{Emergent Time and Diffusion (Heat) Equation\label{DTheat}}

We consider 1 dimensional viscous fluid , and the velocity 
field $\{ u(x); -\infty <x<\infty \}$ describes the velocity distribution in the 1 dim space.  
Let us take the following energy functional\cite{Kikuchi, Kikuchi2} of the velocity-field u(x), 
\bea
I_n[u(x);u_{n-1}(x),\si_{n-1}(x),\rhotil_{n-1}(x)]=             \nn
\intx \{ \frac{\si_{n-1}}{2\rhotil_{n-1}}(\frac{du}{dx})^2 +V(u)+u\frac{dV^1(x)}{dx}
+\frac{1}{2h}(u-u_{n-1})^2\}+I_n^0\ ,\  
V(u)=\frac{m^2}{2}u^2+\frac{\la}{4!}u^4,\nn 
n=1,2,\cdots\com \q
u=u(x)\ ,\ u_{n-1}=u_{n-1}(x)
\ ,\ \si_{n-1}=\si_{n-1}(x)\ ,\ \rhotil_{n-1}=\rhotil_{n-1}(x)
\ ,
\label{DTheat1}
\eea 
where $u_{n-1}(x), \rhotil_{n-1}(x)$ and $\si_{n-1}(x)$ are the step (n-1) distributions of 
the velocity, the {\it mass-density} and the viscosity respectively. 
\footnote{
The 'time'-development term $\Wcal_n(u)$ is generally written as
\bea
I_n[u(x)]=\intx \{ \frac{\si_{n-1}}{2\rhotil_{n-1}}(\frac{du}{dx})^2 +V(u)+u\frac{dV^1(x)}{dx}+\Wcal_n(u)\}
+I_n^0\com\nn
\mbox{where}\q
\Wcal_n(u)=\frac{1}{2h}(u-u_{n-1})^2\q\mbox{or}\q 
\frac{1}{2h^2}(u-2u_{n-1}+u_{n-2})^2
\pr
\label{DTheat1b}
\eea 
The latter case can be treated in the same way and 
is given in App.A. 
Besides the velocity field $u(x)$, the fluid matter density field $\rhotil(x)$ 
and the viscosity field $\si(x)$ generally appears. 

We list the physical dimensions of various quantities.\ \ 
[$x$]=[$l$]=L,\ [$u$]=L/T,\ [$h$]=L/M,\ [$m^2$]=M/L,\ [$\la$]=MT$^2$/L$^3$,\ 
[$\si$]=M$^2$,\ [$\rhotil$] 
=M/L,\ [$I_n$]=ML$^2$T$^{-2}$,\ [$V$]=MLT$^{-2}$,\ [$V^1$]=MLT$^{-1}$. 
Furthermore we have, [$\sqrt{\la\si}$]=(M/L)$^{3/2}$T,\ [$\sqrt{h^3\la\si}$]=
[$m^{-3}\sqrt{\la\si}$]=[$h\sqrt{\la\si}/m$]=T. We use the convention:\ M mass, L length, 
T time. 
             }
$u(x)$ is the general velocity-distribution and becomes $u_n(x)$ by the minimal 
energy principle. 
$I_n^0$ is a 'constant' term which is independent of $u(x)$. Later we will fix it(eq.(\ref{coord4b})). 
$m^2$ is a parameter with the dimension of the mass density:\ (the mass of the fluid-particle)/ 2$l$. 
The quantity $I_n$, (\ref{DTheat1}), has the physical dimension of the energy. 
The {\it velocity potential} $V(u)$ has the mass term and the 4-body interaction term. 
\footnote{
Generally $V(u)$ is chosen for problem by problem. 
The form of $V(u)$ (\ref{DTheat1}) is later, in this text, restricted by the renormalizability condition 
(Sec.4) and this step-flow formulation of the velocity-field theory (Sec.5). 
In the present paper, we take Higgs potential: $m^2 < 0,\ \la>0$. 
As for the coupling parameters
, generally, the 2-body and 4-body couplings depend on $n$:\ $(m^2)_{n-1}$ and 
$\la_{n-1}$. Here we consider the simple case, the couplings do not depend on $n$ 
('time'-independent). See the explanation after eq.(\ref{boltz16}). 
         }
$V^1(x)$ is the (ordinary) position-dependent potential. 
\footnote{
As an example, the gravitational weight force is given by $V^1(x) = g x$. 
}
$\frac{dV^1(x)}{dx}$ is the external source (force) in this velocity-field theory. 
$h$ is a constant which can be regarded as   
the {\it time-separation} for one step. 
$u_{n-1}(x), \si_{n-1}(x)$ and $\rhotil_{n-1}(x)$ are {\it given} distributions 
at the (n-1)-th step evaluation.  The n-th step velocity field  $u_n(x)$  
is given by the {\it minimal principle} of the n-th energy functional $I_n(u)$: 
$\del I_n/\del u(x)=0$ at $u(x)=u_n(x)$. 
This approach is callled "discrete Morse flows method"\cite{Kikuchi, Kikuchi2,MaWitt1203}. 

We may restrict the space region as $-l\leq x\leq l$. 
\footnote{
For the periodic case, $u(x)=u(x+2l), \si(x)=\si(x+2l), \rhotil(x)=\rhotil(x+2l)$ are 
taken. In the text, we consider the general case. 
}
The variation equation $\del I_n(u)=0 (u(x)\ra u(x)+\del u(x))$ gives 
\bea
\frac{1}{h}(u_n(x)-u_{n-1}(x))= \frac{d}{dx}\left(\frac{\si_{n-1}}{\rhotil_{n-1}}\frac{du_n}{dx}\right)
-\frac{\del V(u_n)}{\del u_n}-\frac{dV^1(x)}{dx}\com\nn
\left.\frac{\si_{n-1}}{\rhotil_{n-1}}\frac{du_n}{dx}\right|_{x=-l, l}\ =\ 0,\qqqq
\frac{\del V(u)}{\del u}=m^2u+\frac{\la}{3!}u^3
\com
\label{DTheat3}
\eea 
where we have replaced the {\it minimal solution} by $u_n$. 
From this definition of $u_n(x)$, we have the relation:
\bea
I_n[\un]\q \leq \q I_n[u_{n-1}]
\pr
\label{DTheat3b}
\eea 
See Fig.\ref{EneFunc}. 
From the ordinary (continuous time) experience, we expect
 $I_n[\un]\ \leq\ I_{n-1}[u_{n-1}]$. This relation does not hold. 
Because $I_n[\un]$ is the n-step energy of the system, this 
situation makes the energy calculation more tractable.  

The eq. (\ref{DTheat3})  
describes the n-th step velocity field $u_n(x)$ in terms of $u_{n-1}(x)$ and 
vice versa. Hence it 
can be used for the {\it computer simulation}.
\footnote{ 
The lattice Boltzmann method\cite{succi01} is the most suitable one. 
}

We here introduce the {\it discrete time} variable $t_n$ 
as the step number n of $u_n$. 
\bea
t_n=n h=n \tauzero\times (\frac{h}{\tauzero}) \com\q \tauzero\equiv 
h\sqrt{\la\si_0}/|m|\com\q n=1,\ 2,\ \cdots\nn
t_0\equiv 0 
\com
\label{DTheat4}
\eea 
where $\si_0$ ($>0$) is the representative value (constant) of the system viscosity and 
$\tauzero$ is the time unit. 
\footnote{
Note [$\tauzero$]=T. 
See the footnote of eq.(\ref{DTheat1}). Generally the time $t_n$ can be introduced  
by $t_n =f(n)h$ where $f(n)$ is a function of $n$. The form of $f(n)$ defines 
the time coordinate. The change of the form is the transformation of the time coordinate. 
The simple one is  $f(n)=a n+ b$. In the text 
$f(n)=n$ is taken.  If we take $f(n)=-n$, the time flow is introduced in the inverse 
way. 
              }
It is useful to define the following quantity $A$. 
\bea
A\equiv \sqrt{\la\si_0}/|m|\com\q \mbox{[$A$]=TM/L}\nn
\ttil_n\equiv A\cdot t_n =n\tauzero\com\q \tauzero= A\cdot h
\com\q \mbox{[$\ttil_n$]=[$\tauzero$]=T}\pr
\label{DTheat4b}
\eea 

The eq.(\ref{DTheat3}) is, in terms of the 'renewed' field $u(x, t)$,  expressed as
\bea
\frac{1}{h}(u(x, t_{n-1}+h)-u(x, t_{n-1}))=\nn
\frac{\pl}{\pl x}\left(\frac{\si(x,t_{n-1})}{\rhotil(x,t_{n-1})}\frac{\pl u(x, t_n)}{\pl x}\right) 
-\frac{\del V(u(x, t_n))}{\del u(x, t_n)}
-\frac{\pl V^1(x)}{\pl x}
\com
\label{DTheat5}
\eea 
where we use $u(x, t_n)\equiv u_n(x),\ t_n=t_{n-1}+h$. 
As $h\ \ra\ 0$, we obtain
\bea
\frac{\pl u(x,t)}{\pl t}=\frac{\pl}{\pl x}\left(\frac{\si(x,t)}{\rhotil(x,t)}\frac{\pl u(x,t)}{\pl x}\right) 
-\frac{\del V(u(x, t))}{\del u(x, t)}
-\frac{\pl V^1(x)}{\pl x}
\com
\label{DTheat6}
\eea 
where we have replaced both $t_n$ and $t_{n-1}$ by $t$. 
This is, when  $\si/\rhotil$=const, 
1 dim {\it diffusion} equation with the potential $V(u)$.

We remind that the variational principle for the n-step energy functional $I_n[u(x)]$ 
\  (\ref{DTheat1}), 
$\del I_n=I_n[u+\del u]-I_n[u]=0$, 
gives 
$u_n(x)$ for the given $u_{n-1}(x), \si_{n-1}(x)$ and $\rhotil_{n-1}(x)$. 
We regard the increase of the step number as the {\it time development}. 
\footnote{
Time is defined here by the {\it energy-minimal principle}. 
             }
Taking into account the fact that, at the (n-1)th-step, the matter-particle 
at the point $x$ flows at the speed of $u_{n-1}(x)$, 
the energy functional $I_n$, (\ref{DTheat1}), 
should be replaced by the following one\cite{Kikuchi, Kikuchi2}.

\bea
\Itil_n[u(x)]=\intx \{ \frac{\si_{n-1}}{2\rhotil_{n-1}}(\frac{du}{dx})^2 +V(u)+u\frac{dV^1(x)}{dx}\nn
+\frac{1}{2h}(u(x+Ah~u_{n-1})-u_{n-1})^2\}+\Itil
_n^0\com\nn 
V(u)=\frac{m^2}{2}u^2+\frac{\la}{4!}u^4, \q 
n=1,2,\cdots\ ,\nn 
u=u(x)\ ,\ u_{n-1}=u_{n-1}(x)
\ ,\ \si_{n-1}=\si_{n-1}(x)\ ,\ \rhotil_{n-1}=\rhotil_{n-1}(x)
\ .
\label{DTheat6b}
\eea 
Note that $u(x)-u_{n-1}(x)$ in eq.(\ref{DTheat1}) is replaced by $u(x+Ah~u_{n-1}(x))-u_{n-1}(x)$. 
($\tauzero=Ah$ is the time difference between 1 step. )
For the simple case of {\it no potential},\ $V=0$\ , and 
no external force,\ $\frac{dV^1}{dx}=0$, 
\bea
J_n[u(x)]=\intx \{\frac{\si_{n-1}}{2\rhotil_{n-1}}(\frac{du}{dx})^2 +\frac{1}{2h}(u(x+Ah~u_{n-1}(x))-u_{n-1}(x))^2\}
+'\mbox{const}',
\label{DTheat7}
\eea 
The above functional is equivalent to $I_n[u(x)]$ with the potential.  
\bea
V(u)=A(u(x)-u_{n-1}(x))u_{n-1}(x)\frac{du(x)}{dx}+O(h)
\pr
\label{DTheat8}
\eea 
where we consider the case of {\it sufficiently-small} $h$. Eq.(\ref{DTheat3}) gives us 
the following equation as the minimal equation for $J_n[u]:\ \del J_n[u]=0$
\footnote{
\bea
\del J_n=\intx \{\frac{\si_{n-1}}{\rhotil_{n-1}}\frac{du}{dx}\frac{d}{dx}(\del u) +
\frac{1}{h}(u-u_{n-1})\del u +\del u\cdot Au_{n-1}\frac{du}{dx} +
Au_{n-1}(u-u_{n-1})\frac{d}{dx}\del u  \}
\com
\label{DTheat9c}
\eea 
The first and last terms give the boundary terms in (\ref{DTheat9}). 
            }
\bea
\frac{1}{h}(u_n(x)-u_{n-1}(x))=  \frac{d}{dx}X[u_n(x),u_{n-1}(x)]
-A~u_{n-1}(x)\frac{du_n(x)}{dx} \com\nn
X[u_n(x),u_{n-1}(x)]|_{x=-l, l}\ =\ 0\ ,\nn  
X[u_n(x),u_{n-1}(x)]=A (u_n(x)-u_{n-1}(x))u_{n-1}(x)+\frac{\si_{n-1}}{\rhotil_{n-1}}\frac{du_n}{dx}
\ .
\label{DTheat9}
\eea 
Hence the step-wise {\it recursion relation} (\ref{DTheat3}) is corrected as
\bea
\mbox{[Comp 1]}\q u_n(x)\ \mbox{Equation}\qqqq\qqqq\qqqq\qqqq  \nn
\frac{1}{h}(u_n(x)-u_{n-1}(x))+A~u_{n-1}(x)\frac{d\un(x)}{dx} \nn
=\frac{d}{dx}\left(
X[u_n(x),u_{n-1}(x)]
                 \right) 
-\frac{\del V(u_n)}{\del u_n}-\frac{dV^1(x)}{dx}\ ,\nn
X[u_n(x),u_{n-1}(x)]|_{x=-l, l}\ =\ 0
\pr
\label{DTheat9b}
\eea 
This equation is obtained by $(\del\Itil_n[u]/\del u)|_{u_n}=0$. 

  As done before, let us replace the step number $n$ by the {\it discrete} time $t_n=nh$. 
Taking the continuous time limit ($\tauzero=Ah\ \ra\ 0$), we obtain 
\bea
A\frac{\pl u(x,t)}{\pl \ttil}+Au(x,t)\frac{\pl u(x, t)}{\pl x} =
\frac{\pl}{\pl x}\left(\frac{\si(x,t)}{\rhotil(x,t)}\frac{\pl u(x,t)}{\pl x}\right) 
-\frac{\del V(u(x, t))}{\del u(x, t)}
-\frac{\pl V^1(x)}{\pl x}\ ,\nn
\left.\frac{\si(x,t)}{\rhotil(x,t)}\frac{\pl u(x,t)}{\pl x}\right|_{x=-l, l} =\ 0\com
\label{DTheat10}
\eea 
where two continuous times ($t$ and $\ttil=At$) are used: 
$t_n=nh\ra t$ and $\ttil_n=nAh=n\tauzero\ra \ttil (=A~t)$ as $n\ra \infty$. 
\footnote{
The higher-order terms, $- A^2h\frac{\pl}{\pl x}(\frac{\pl u(x,t)}{\pl \ttil}u(x,t))$ in the LHS of the first equation and 
$ A h \frac{\pl u(x,t)}{\pl \ttil}u(x,t)$ in the LHS of the last equation
are ignored. 
             }
This is
, when 
$\si/\rhotil$=const, 
{\it Burgers equation} (with the velocity potential $V(u)$ and the external 
force $\frac{\pl V^1}{\pl x}$) and is considered to be 1 dimensional 
Navier-Stokes equation. 
Note that the non-linear term in the LHS of eq.(\ref{DTheat10}) appears 
not from the potential ( velocity-field interaction) but from 
the change $u(x)$ in (\ref{DTheat1}) to $u(x+Ahu_{n-1})$ in (\ref{DTheat6b}), 
namely, the consistency between the (space) {\it coordinate} $x$ and the {\it velocity} 
field $u(x)$ in the step-flow.   
The differential operator: 
$\frac{\pl}{\pl \ttil}+u\frac{\pl}{\pl x}\equiv\frac{D}{D \ttil}$, appearing in LHS, is called 
{\it Lagrange derivative}. 
The relations between $\si_{n-1}$ and $\si_{n}$, and $\rhotil_{n-1}$ and $\rhotil_{n}$, 
which describe their step-flow ('time'-development), 
are given in Sec.\ref{boltz}. 

The equation (\ref{DTheat10}), for the massless case $m=0$, is invariant under 
the {\it global scale transformation}. 
\bea
t\q\ra\q \e^{2\vep}t\com\q x\q\ra\q \e^\vep x\com\nn
\frac{\pl}{\pl x}=\pl_x\q\ra\q\e^{-\vep}\pl_x\com\q
\frac{\pl}{\pl t}=\pl_t\q\ra\q\e^{-2\vep}\pl_t\com\nn
u(x,t)\q\ra\q \e^{-\vep} u(\e^\vep x, \e^{2\vep}t)\com\q
\frac{\si(x,t)}{\rhotil(x,t)} \q\ra\q \frac{\si(\e^\vep x, \e^{2\vep} t)}{\rhotil( \e^\vep x,\e^{2\vep} t)}
\com\nn
V^1(x)\ \ra\ \e^{-2\vep}V^1(\e^\vep x)\ ,\ 
V(u(x,t))\ \ra\ \e^{-2\vep}V( u(\e^\vep x, \e^{2\vep}t)  )\ (\la\ \ra\ \e^{-2\vep}\la)
\com
\label{DTheat11}
\eea 
where $\vep$ is the real constant parameter. 
\footnote{
If we consider the mass $m$ here appears in some dynamical way 
(, for example, through the spontaneous breakdown ), $m\ra\e^{-\vep} m$ 
makes eq.(\ref{DTheat10}) 
global scale invariant. 
}
When it is small :\ $|\vep|\ll 1$, the variation $\del u$ is given by 
\bea
\del u=
\vep \{ -u+x~\pl_xu+2t~\pl_tu\}+O(\vep^2)
\com
\label{DTheat12}
\eea 
 
For simplicity, we explain in one space-dimension (dim). The generalization to 2 dim 
and 3 dim is straightforward. Furthermore the ordinary field theory (not using the velocity field 
but the particle field) is described by this step-wise approach in App.A.

\vspace{5mm}
\section{Space Orbit (Path) and Total Energy\label{coord}}

 The space coordinate $x$ always appears in the velocity field $u_n(x)$. 
We can introduce the n-th step space coordinate $x_n$ as
\bea
u_n(x_n)=\frac{x_{n+1}-x_{n}}{\tauzero}\com\q n=0,1,2,\cdots\nn
x_{n+1}=x_n+\tauzero u_n(x_n)
\com
\label{coord1}
\eea 
where $x_0$ is a given initial position and $\tauzero= h\sqrt{\la\si_0}/|m|=hA$ in eq.(\ref{DTheat4}). We can trace the position of 
the matter-point, which is at $x_0$ as the initial (0-th) step, along 
the step process:\ $x_0, x_1, x_2, \cdots$. After N steps, the matter-point 
reaches the following point. 
\bea
x_N=x_{N-1}+\tauzero u_{N-1}(x_{N-1})=x_0+\tauzero \sum_{n=0}^{N-1}u_n(x_n)\pr
\label{coord2}
\eea 
In terms of the continuous time $\ttil=At$, 
\bea
x(T)=x_0+\int_{0}^{T}u(x(t),t)d\ttil \com
\label{coord3}
\eea 
where $T=N~Ah, \ttil_n=n~Ah=At_n, x(t_n)=x_n, u(x,t_n)=u_n(x)$. $x(t)$ is the orbit or path of the 
matter-point, at $x_0$ initially,  "moving" in the N steps.  

At the n-th step, the total energy of the system, $E_n$, is given by
\bea
E_n\equiv \Itil_n[u_n]\com\q
\mbox{where}\qqqqq 
\left.\frac{\del \Itil_n[u]}{\del u(x)}\right|_{u=u_n}=0\com \nn 
\Itil_n[u]=\intx \{ \frac{\ka_{n-1}}{2}(\frac{du}{dx})^2 +V(u)+u\frac{dV^1}{dx}+
\frac{1}{2h}(u(x+hAu_{n-1})-u_{n-1})^2\}+\Itil_n^0
\ ,\nn
\ka_n(x)\equiv\frac{\si_n(x)}{\rhotil_n(x)}\com\qq
u=u(x)\com\q u_{n-1}=u_{n-1}(x)\com
\label{coord4}
\eea 
where $\ka_n=\ka_n(x)$ is introduced. $\Itil_n^0$ is taken as
\bea
\Itil_n^0=-\intx \{ u_n\frac{dV^1}{dx}+
\frac{1}{2h}(u_n(x+hAu_{n-1})-u_{n-1})^2\}\nn
+\intx \{ \frac{\ka_0}{2}(\frac{du_0}{dx})^2 +V(u_0)+u_0\frac{dV^1}{dx}
+\frac{\rhotil_0}{2}{u_0}^2\}
\pr
\label{coord4b}
\eea 
The above one ($\Itil_n^0$) is chosen in such a way that the total energy $E_n=\Itil_n[u_n]$ keeps 
the initial energy (the second integral of (\ref{coord4b})) when the {\it dissipative} 
terms, $\frac{\ka_{n-1}}{2}(\frac{du}{dx})^2$ and $V(u)$, do not appear. $\Itil_n[u_n]$ 
is the n-th step dissipative energy plus the initial energy. 

The system total energy $E_n$ generally {\it changes} as the step number, n,  increases. 
\bea
W(t_n)\equiv\frac{dE(t_n)}{d\ttil_n}=
\frac{h}{\tauzero}\frac{dE(t_n)}{dt_n}\equiv\frac{\Itil_{n+1}[u_{n+1}]-\Itil_n[u_n]}{\tauzero}=\nn
\frac{1}{\tauzero}\intx \left\{ \half\left( \ka_n (\frac{du_{n+1}}{dx})^2 -\ka_{n-1} (\frac{du_n}{dx})^2\right) +V(u_{n+1})-V(u_n)+(u_{n+1}-u_n)\frac{dV^1}{dx} \right.
\nn
\left.+\frac{1}{2h}\{  (u_{n+1}+hAu_n\frac{du_{n+1}}{dx}-u_n)^2 -(u_n+hAu_{n-1}\frac{du_n}{dx}-u_{n-1})^2\}
\right\}
+\frac{\Itil_{n+1}^0-\Itil_n^0}{\tauzero}
,
\label{coord5}
\eea 
where $E(t_n)\equiv E_n$ and $W(t_n)$ is the energy rate.  From the above 
formula we get the expression for the energy at $t=Nh=t_N$. 
\bea
E(t_N)=\Itil_N[u_N]=\Itil_{N-1}[u_{N-1}]+\tauzero W(t_{N-1})=\Itil_1[u_1]+\tauzero \sum_{n=1}^{N-1}W(t_n)
\pr
\label{coord6}
\eea 

When we regard the process of the increasing step-number as 
the time development, the system generally does {\it not} conserve energy. 
\footnote{
We will numerically confirm the non-conservation later in Sec.\ref{qm}. 
             }
$E(t_n)$($=\Itil_n[u_n]$) generally changes step by step. 
We can physically understand that 
the increase or decrease of the total system energy is given or subtracted by the {\it outside} ({\it environment}). 
The energy functional (\ref{DTheat6b}) describes the {\it open-system} dynamics. 
When $E(t_n)$ satisfies 
\bea
W(t_n)=\frac{h}{\tauzero}\frac{dE(t_n)}{dt_n}\q\ra\q W_0(\mbox{constant})\q\mbox{as}\ n\ra\infty
\com
\label{coord6b}
\eea 
we say the system finally reaches the {\it steady} energy-state. 
\footnote{
Energy {\it constantly} comes in or goes out. App.C shows such an example. 
} 
For the special case of $W_0=0$, 
we say the system finally reaches the {\it constant} energy-state
\footnote{
Energy does not go out and does not come in. 
} .

In Sec.\ref{qm}, we treat the $W_0=0$ case. As the example of the more general case 
($W_0\neq 0$), another model is given in App.C.

\section{Statistical Fluctuation Effect\label{fluct}}

We are considering the system of {\it large number of matter-particles}, 
hence the physical quantities, such as energy and entropy, are given by the 
{\it statistical average}. In the present approach, the system behavior $u_n(x)=u(x,t_n)$ is completely 
determined by eq.(\ref{DTheat9b}) when the initial configuration $u_0(x)=u(x,0)$ is given. 
We have obtained the solution by the {\it continuous} variation $\del u(x)$ to $\Itil_n[u]$, (\ref{DTheat6b}). 
In this sense, $u_n(x)$ is the 'classical path'. 
Here we should note that the present formalism is 
an {\it effective} approach to calculate the physical properties of this {\it statistical} system. 
{\it Approximation} is made in the following points: 
\begin{description}
\item[1)]\ \ 
So far as $h\neq 0$, the {\it finite time-increment} gives {\it uncertainty} 
to  the minimal solution $u_n(x)$. 
This is because we cannot specify the minimum configuration definitely, 
but can only do it with {\it finite} uncertainty.  
\item[2)]\ \ 
We do not measure the initial $u_0(x)$. Hence the velocity distribution $u_n(x)$ 
fluctuates due to the initial condition ambiguity.  
\item[3)]\ \ 
The real fluid matter is made of many micro particles with small but {\it non-zero size}. The existence 
of the characteristic particle size gives uncertainty to the minimal solution in this (space-)continuum formalism. 
Furthermore the particle size is not constant but does distribute in the statistical way. The {\it shape} of 
each particle differs. The present continuum formalism has limitation to describe the real situation 
accurately.   
\item[4)]\ \  
\ The system energy generally changes step by step. 
The present model (\ref{DTheat6b}) describes an {\it open-system}. 
It means the present system {\it energetically} 
interacts with the outside. Such interaction is caused by the 
dissipative term in (\ref{DTheat6b}). 
\end{description}
We claim the fluctuation, in the present approach, 
comes {\it not} from the {\it quantum effect} but from the {\it statistics} due to 
the 
uncertainty which comes from 
the {\it finite} time-separation and the spacial distribution of {\it size} and {\it shape} 
of the content particles. 

To take into account this fluctuation effect, we {\it newly} define the n-th energy functional 
$\Ga_n[u(x)]=\Ga_n[u(x);u_{n-1}(x),\si_{n-1}(x),\rhotil_{n-1}(x)]$ 
in terms of the original one $\Itil_n[u(x)]$, 
(\ref{DTheat6b}), using the {\it path-integral}(in the velocity-field).
\footnote{
$u(x)$ is a velocity distribution over the space $x$. Hence this 
path-integral is the integral (sum) of all possible {\it distributions}. 
} 
\bea
\e^{-\alinv\Ga_n[u_n(x)]}=\int\left.\Dcal q(x)\e^{-\alinv \Itil_n[u(x)]}\right|_{u=u_n+\sqrt{\al}q}
                                                               \com\qqqqq\qqq\nn
\Itil_n[u(x)]=\intx \{ \frac{\ka_{n-1}}{2}(\frac{du}{dx})^2 +V(u)
+u(x)\frac{dV^1(x)}{dx}  \nn
+\frac{1}{2h}(u(x+h A u_{n-1})-u_{n-1})^2\}+\Itil^0_n,\nn 
V(u)=\frac{m^2}{2}u^2+\frac{\la}{4!}u^4
\com\q    \left.\frac{\del \Itil_n[u]}{\del u(x)}\right|_{u=u_n}=0\pr   \qqqqq
\label{fluct1}
\eea 
In the above path-integral expression, {\it all} paths(distributions) $\{ u(x); -l\leq x\leq l \}$ are taken into account. 

We are considering the minimal path $u_n(x)$ as the dominant configuration 
and the small deviation $q(x)$ around it. 
\bea
u(x)=u_n(x)+ \sqral q(x)\com\q |\sqral q|\ll |u_n|\com\q 
\left.\frac{\del \Itil_n[u]}{\del u}\right|_{u=u_n}=0\com
\label{fluct2}
\eea 
In eq.(\ref{fluct1}) and eq.(\ref{fluct2}), a {\it new} expansion parameter $\al$ is introduced. \nl
([$\al$]=[$I_n$]=ML$^2$T$^{-2}$, $[q]=1/\sqrt{M}$) 
As the above formula shows, 
$\al$ should be small. The concrete form should be chosen depending on problem 
by problem. It should {\it not} include Planck constant, $\hbar$, because the fluctuation does not 
come from the quantum effect. Hence the parameter $\al$ should be chosen as
\begin{description}
\item[1)]the dimension is consistent,
\item[2)]it should have the small scale parameter which characterizes 
the relevant physical phenomena such as the mean-free path of the fluid particle, 
\item[3)]the precise value should be best-fitted with the experimental data. 
\end{description}

  The background-field method\cite{DeW67, tHooft73} tells us to do the Taylor-expansion around $\un$. 
\footnote{
The background-field method was originally introduced to {\it quantize} the gravitational field theory. The geometrical viewpoint is introduced here in the former treatment of the fluid matter. 
We borrow the method only to define the {\it statistical} distribution measure. The present 'splitting' is 
$u=u_n+\sqral q$,\ not $u=u_n+\sqrt{\hbar} q$. 
}
\bea
\Itil_n[u(x)]=\Itil_n[u_n(x)+\sqral q(x)]=\sum_{l=0}^{\infty}\al^{l/2}\frac{q(x)^l}{l!}\left.\frac{\del^l\Itil_n[u]}{\del u(x)^l}\right|_{u_n}
=\sum_{l=0}^{\infty}S_l[u_n]
\com
\label{fluct2b}
\eea 
Then the n-th energy functional, $\Ga_n[u(x)]$, is expressed in the perturbed form 
up to the second order (w.r.t. $\sqral$). 
\bea
\e^{-\alinv \Ga_n[u_n(x)]}=
\int\Dcal q(x)\exp (-\alinv (S_0+S_2 +O(q^3)))=\e^{-\alinv \Itil_n[u_n(x)]}\times\nn
\int\Dcal q\exp\left[\int dx
\left\{
-\frac{q(x)^2}{2!}(\frac{\del}{\del u(x)})^2\Itil_n[u]|_{u_n}+O(q^3)
\right\}
                      \right]\com\q
\frac{\del}{\del u(x)}\Itil_n[u]|_{u_n}=0\com                                \nn
S_0=\Itil_n[u_n]\com\q S_1=\int dx q(x)\left.\frac{\del \Itil_n[u]}{\del u}\right|_{u_n}=0\com\nn
\alinv S_2=
\frac{q(x)^2}{2!}(\frac{\del}{\del u(x)})^2\Itil_n[u]|_{u_n}=
\int dx\left[ \frac{\ka_{n-1}}{2}(\frac{dq}{dx})^2+\right.\nn
\left. (\frac{m^2}{2}+\frac{\la}{2}{u_n}^2)q^2+\frac{1}{2h}(q+h A u_{n-1}\frac{dq}{dx})^2\right]\nn
\equiv \int dx\left[ \half \frac{d}{dx}(\ka_{n-1}q\frac{dq}{dx})+\half q D q+O(h)\right]\com  \nn 
D\equiv -\frac{d}{dx}\ka_{n-1}\frac{d}{dx}
+\la {u_n}^2+m^2+\frac{1}{h}+2u_{n-1}A\frac{d}{dx}
\com
\label{fluct3}
\eea 
where we make the Gaussian(quadratic, 1-loop) approximation. 
\footnote{
$O(h)=(h/2)u_{n-1}^2(dq/dx)^2$ may be ignored for $h\ll 1$. 
}
\bea
\e^{-\alinv \Ga_n[u_n(x)]}=\e^{-\alinv \Itil_n[u_n(x)]}\times(\det D)^{-1/2}\com\nn
(\det D)^{-1/2}=\exp\left\{ -\half \Tr\ln D \right\}
=               \exp
\left\{
\half\Tr\int_0^\infty\frac{\e^{-\tau D}}{\tau}d\tau +\mbox{const}
\right\}
                      \com
\label{fluct4}
\eea 
where $\tau$ is called Schwinger's proper time\cite{Schwinger51}. ([$\tau$]=[$D^{-1}$]=L/M.)

To rigorously define the inside of the above exponent, we introduce Dirac's 
abstract state vector $|x>$ and $<x|$. 
\bea
<x|\e^{-\tau D}|y>\equiv G(x,y;\tau)\com\nn
(\frac{\pl}{\pl\tau}+D)G(x,y;\tau)=0\com\q \lim_{\tau\ra +0}G(x,y;\tau)=\del(x-y)\com\nn
D=-\frac{d}{dx}\ka_{n-1}(x)\frac{d}{dx} + 2u_{n-1}A\frac{d}{dx} -\Vbar(x)\com\q 
\Vbar(x)=-\la {u_n(x)}^2-m^2-\frac{1}{h}   
\pr
\label{fluct5}
\eea 
$G(x,y;\tau)$ is called {\it heat-kernel}\cite{Schwinger51}. 

In App.B, we evaluate, for the case $\ka_{n-1}=\ep^{-1}$(const) and $A=0$, 
$\ln(\det D)^{-1/2}=\half\int_0^\infty d\tau \Tr G(x,y)/\tau=  
\half\int_0^\infty d\tau \int_{-l}^{l}dx~ G(x,x)/\tau$. Up to the first order of $\Vbar$, 
the result is given by
\bea
\frac{l}{\sqrt{\pi}}\sqrt{\ep\La}-\frac{1}{2\sqrt{\pi\ep\mu}}\int_{-l}^l dz \ep (\la\un(z)^2+m^2+\frac{1}{h})
                      \com
\label{fluct6}
\eea 
where 
the {\it infrared cut-off} parameter $\mu\equiv \sqrt{\si_0}/l$ and  
the {\it ultraviolet cut-off} parameter $\La\equiv h^{-1}$ are introduced. 
$\ep^{-1}\equiv \si_0/\rhotil_0 =1$. 
We see the mass parameter $m^2$ {\it shifts} under the influence of the fluctuation. 
\footnote{
This corresponds to renormalization of "mass" $m^2$ in the field theory. 
When natural cut-offs (IR and UV) are there in the system model-parameters, 
the divergences coming from the space integral and the mode summation 
are effectively expressed by "large" but finite quantities. 
}
\bea
m^2\q\ra\q m^2+\frac{\al}{\sqrt{\pi\ep\mu}}\ep\la=m^2+\al\la\sqrt{\frac{l\rhotil_0}{\pi\si_0\sqrt{\si_0}}}
                      \pr
\label{fluct7}
\eea 
And the bottom of the potential shifts as 
\bea
V(u_{min})\q\ra\q V(u_{min})+\al\left\{ \half\sqrt{\frac{\ep\La}{\pi}}-\frac{\ep}{2\sqrt{\pi\ep\mu}}(m^2+\frac{1}{h}) \right\}\nn
=V(u_{min})+\al\sqrt{\frac{\rhotil_0}{4\pi\si_0}}\left\{ \frac{1}{\sqrt{h}}-\sqrt{ \frac{l}{\sqrt{\si_0}}}(m^2+\frac{1}{h})  \right\}
\pr
\label{fluct8}
\eea 

 The coupling $\la$ is also shifted by the O($\Vbar^2$) correction. 
\footnote{
The coupling ($\la$) shift can be obtained from $O(\Vbar)$ result (\ref{fluct8})
by assuming the "renormalization" consistently works. Noting 
$V(u_{min})=-6m^4/\la$, $\la$ should shift as 
$$
\la\q\ra\q \la+\frac{\al}{6}(\frac{\la}{m^2})^2
\sqrt{\frac{\rhotil_0}{4\pi\si_0}}\frac{1}{\sqrt{h}}\left\{1-\sqrt{ \frac{l}{\sqrt{\si_0}h}}\right\}
$$
}
The shift of these parameters 
corresponds to the {\it renormalization} in the field theory\cite{Wein95}. 
In this effective approach, we have {\it physical} 
cut-offs $\mu$ and $\La$ which are expressed by the (finite) parameters appearing in the starting 
energy-functional. When the functional (\ref{DTheat6b})  (effectively) 
works well, all effects of the statistical fluctuation reduces to the simple shift of the original parameters. 
This corresponds to the {\it renormalizable} case in the field theory. 
We consider this case in the following.

\section{Boltzmann's Transport Equation\label{boltz}}

\subsection{Distribution function and Boltzmann's transport equation}
We use, for simplicity, the original names for the shifted parameters. 
The step-wise development equation (\ref{DTheat9b}),  with 
$\del V/\del u=m^2u+\frac{\la}{3!}{u}^3+u_{n-1}\frac{d\un}{dx}$ and $V^1=0$, 
is written as
\bea
              \mbox{[Comp 1']}\q u_n(x)\ \mbox{Equation}\qqqq\qqqq  \nn
\frac{1}{h}(u_n(x)-u_{n-1}(x))=\frac{d}{dx}\left(\ka_{n-1}\frac{d u_n}{dx}
+A (u_n(x)-u_{n-1}(x)) u_{n-1}(x)
\right) \nn
-m^2u_n-\frac{\la}{3!}{u_n}^3-A~u_{n-1}\frac{du_n}{dx}\com\q
X[u_n(x),u_{n-1}(x)]|_{x=-l, l}\ =\ 0\com\nn 
X[u_n(x),u_{n-1}(x)]=A (u_n(x)-u_{n-1}(x))u_{n-1}(x)+\frac{\si_{n-1}}{\rhotil_{n-1}}\frac{du_n}{dx}
\com
\label{boltz1}
\eea 
where $  \ka_n(x)=\si_n(x)/\rhotil_n(x)$(eq.(\ref{coord4})).
When the system reaches the {\it equilibrium state} after sufficient recursive computation ($n\gg 1$), 
we may assume $u_{n-1}(x)=u_n(x)\equiv \uinf (x), 
\ka_{n-1}(x)=\si_{n-1}(x)/ \rhotil_{n-1}(x)=\ka^\infty$(const.). 
$\uinf (x)$ satisfies
\bea
\ka^\infty\frac{d^2\uinf}{dx^2}-m^2\uinf-\frac{\la}{3!}({\uinf})^3-A~\uinf\frac{d\uinf}{dx}=0\com\q
\ka^\infty\equiv \frac{\si^\infty}{\rhotil^\infty}
\pr
\label{boltz2}
\eea 
\footnote{
When $\ka^\infty=1$, 
a solution is $\uinf=\sqrt{(-3!/\la)m^2}$ (constant) for $m^2<0$. 
$V(u)$ is Higgs potential in the velocity-space. 
}

Here we introduce the {\it particle number} density, $\rho_n(x)$. 
\footnote{
Here we list the physical dimension of some quantities appearing in this section. 
[$x$]=L, [$\un$]=[$v$]=L/T, [$\kB \Tcal$]=ML$^2$/T$^2$, 
[$\rho_n(x)$]=L$^{-1}$, [$\rhotil_n(x)$]=ML$^{-1}$, [$f_n$]=T/L$^2$, 
[$P_n$]=ML/T$^2$, [$q_n$]=ML$^2$/T$^3$. 
}
The {\it continuity equation} is given by
\bea
\mbox{[Comp 2]}\q \rho_n(x)\ \mbox{Equation}\qqqq\qqqq  \nn
\frac{1}{h}(\rho_n(x)-\rho_{n-1}(x))+\frac{d}{dx}(\rho_{n-1}(x)u_{n-1}(x))=0
\pr
\label{boltz4b}
\eea 
This relation defines the step-flow of $\rhon(x)$. 

The {\it distribution function} $f_n(x, v)$ is introduced in the following way.  
The probability for the matter particle in the space interval $x\sim x+dx$ 
and the velocity interval $v\sim v+dv$, at the step $n$, is given by
\bea
\frac{1}{\Nbar_n} f_n(x, v) dx dv
\com\q
\Nbar_n\equiv \int dx dv f_n(x,v)\com
\label{boltz3}
\eea 
where $\Nbar_n$ is the total particle number of the system at the step $n$.\cite{BMB04} 
Then the n-th {\it distribution} $\fn (x,v)$ and 
the {\it equilibrium distribution} $f^\infty(x,v)$ are expressed as 
\bea
\un(x)=\frac{1}{\rho_n(x)}\int v \fn(x, v) dv\com\q
\rho_n(x)=\int dv f_n(x,v), \nn
\uinf(x)=\frac{1}{\rho_\infty(x)}\int v \finf(x, v) dv\com\qqqq\qqqq\nn 
\un(x)\ra\uinf(x)\q\mbox{and}\q \fn(x,v)\ra\finf(x,v)\q\mbox{as}\q n\ra\infty
\com
\label{boltz4}
\eea 
where $\uinf(x)$ is the {\it equilibrium} velocity distribution. 
The expression $u_n(x)$ in eq.(\ref{boltz4}) guarantees the {\it momentum conservation} at each point, $x$. 
\bea
0=\rhotil_n(x)\int dv (v-\un(x))\fn(x,v)
\pr
\label{boltz7}
\eea 

The recursion relation (\ref{boltz1}) is expressed, in terms of the distribution functions, as
\bea
\frac{1}{h}\left[
\ftil_n(x+hA~u_{n-1}(x), v)-\ftil_{n-1}(x, v)
\right]=\nn
\frac{\pl}{\pl x}\left(\ka_{n-1}(x)\frac{\pl \ftil_n(x,v)}{\pl x}\right)
-m^2\ftil_n(x,v)
-\frac{\la}{3!}\ftil_n (x, v){\un(x)}^2\com\nn
\ftil_n(x,v)\equiv\frac{\fn (x, v)}{\rho_n(x)},\ 
\un(x)=\frac{1}{\rho_n(x)}\int v \fn(x, v) dv,\ 
\ka_{n-1}(x)=\frac{\si_{n-1}(x)}{\rhotil_{n-1}(x)}\ .
\label{boltz5}
\eea 
This is the {\it Boltzmann's transport equation} for the 2-body and 4-body 
velocity-interactions (Higgs-type velocity potential)\cite{BMB04}. 
We can express the step-wise expression (\ref{boltz5}) in the continuous time $t$ 
form as in Sec.\ref{DTheat}. 
This is the integro-differential equation for $f_n(x,v)$. 
The right hand side (RHS) of the top equation of (\ref{boltz5}) is called the {\it collision term}. 
\footnote{
The lattice Boltzmann method\cite{succi01} is the computer algorithm to determine the 
distribution function $f_n(x,v)$ using Boltzmann's transport equation (\ref{boltz5}). 
             }

We now introduce some physical quantities used in 
the {\it non-equilibrium statistical mechanics}. 
The {\it entropy} $S_n$ and the {\it total particle-number} $\Nbar_n$ are defined by
\bea
S_n\equiv -\kB\int dv\int dx f_n(x,v)\ln f_n(x,v)\com\nn
\Nbar_n=\int dx \rho_n(x)=\int dx\int dv f_n(x,v)\com
\label{boltz6}
\eea 
where $\kB$ is Boltzmann's constant. Another physical quantities will be presented. 

Besides the particle-number density $\rho_n(x)$, 
we already have introduced the {\it mass} density $\rhotil_n(x)$ at step $n$. 
We here consider the case of  one kind particle. 
\bea
\frac{\rhotil_n(x)}{\rhon(x)}=m_1\ (\ \mbox{constant} \ )
\com
\label{boltz8}
\eea 
where $m_1$ is the particle mass. 
\footnote{
For $k$ kinds particles, we introduce $\rhon$ and $\rhotil_n$ fields for 
each one:\ $\{ \rhon^i, \rhotil_n^i \ | i= 1,2,\cdots , k \}$ with the relation 
${\rhotil_n}^i (x)/\rhon^i (x)=m_i$. 
            }
In this case the total mass at the step $n$, $M_n$, is given by 
\bea
M_n\equiv \int dx~ \rhotil_n(x)=m_1\int dx~ \rho_n(x)=m_1\Nbar_n
\pr
\label{boltz9}
\eea 
The {\it temperature} distribution $\Tcal_n(x)$, the {\it heat current} distribution $q_n(x)$ and 
the {\it pressure} $P_n(x)$ are given by 
\bea
\frac{1}{2}\kB \Tcal_n(x)\equiv\frac{1}{\rho_n(x)}  \int dv \frac{m_1}{2}(v-\un(x))^2\fn(x,v)\com\nn
q_n(x)\equiv \int dv \frac{m_1}{2}(v-\un(x))^3\fn(x,v)\com\nn
P_n(x)\equiv m_1 \int dv (v-\un(x))^2\fn(x,v)=\kB\rho_n(x)\Tcal_n(x)
\com
\label{boltz10}
\eea 
where $\kB$ is Boltzmann's constant. 
\footnote{
The last equation $P_n(x)=\kB\rho_n(x)\Tcal_n(x)$ is the {\it equation of state}, which is here 
 valid by their definition. 
}

In this subsection, $\Tcal_n(x), q_n(x), P_n(x)$ and $S_n$ are introduced 
using the distribution function $f_n(x,v)$. Another definition is given in the next 
subsection.  
\subsection{Viscosity, heat conductivity and renormalization}
Using the transport equation (\ref{boltz5}) which $f_n(x,v)$ satisfies, we expect the following  
two equations are satisfied. 
The first one is the rephrasing of the Navier-Stokes equation (\ref{boltz1}). 
\bea
\mbox{[Comp 3]}\q P_n(x) \mbox{ or } \Tcal_n(x)\ \mbox{Equation}\qqqq\qqqq  \nn
\frac{1}{h}(u_n(x)-u_{n-1}(x))+A~u_{n-1}\frac{du_n}{dx}=-\frac{1}{\rhotil_n(x)}\frac{dP_n(x)}{dx}
\pr
\label{boltz11}
\eea 
The second one is the energy equation. 
\bea
\mbox{[Comp 4]}\q q_n(x)\ \mbox{Equation}\qqqq\qqqq  \nn
\half\rhon(x) \kB A\{ \frac{1}{h\cdot A}(\Tcal_n(x)-\Tcal_{n-1}(x))+u_{n-1}\frac{d\Tcal_n}{dx}\}=
-\frac{dq_n(x)}{dx}-P_n(x)\frac{d\un(x)}{dx}
\com
\label{boltz12}
\eea 
where $\Tcal_n = P_n/\kB\rho_n$ is defined in (\ref{boltz11}). 
We explain how the above two relations, (\ref{boltz11}) and (\ref{boltz12}), 
are valid. 

We define n-th viscosity $\si_n(x)$ by the following equation.
\bea
\mbox{[Comp 5]}\q \si_n(x)\ \mbox{Equation}\qqqq\qqqq  \nn
P_n(x)=-\si_n(x)\frac{du_n}{dx}
\pr
\label{boltz13}
\eea 
This gives the step-flow of $\si_n(x)$ because other two components ($P_n$ and $du_n/dx$) are 
already determined by $\si_{n-1}(x)$. 
\footnote{
When 
$\si_n(x)$, defined by eq.(\ref{boltz13}), is expressed as $\si_n(x)=F(du_n/dx)$, 
the present fluid has different names depending on the form of F. \nl
1)\ F=$\mu$(const),\ Newtonian;\ \ 
2)\ F$\propto (du_n/dx)^{n-1}\ (n>1)$,\ \ Dilatant;\ \ 
3)\ F$\propto (du_n/dx)^{n-1}\ (n<1)$,\ \ Quasi-viscous. 
In the present treatment, both $P_n(x)$ and $du_n/dx$ are already obtained 
at this stage, so the form of $F(du_n/dx)$ is here derived, not assumed. 
}
Then the RHS of eq.(\ref{boltz11}) is written as
\bea
-\frac{1}{\rhotil_n(x)}\frac{dP_n(x)}{dx}=\ka_n(x)\frac{d^2u_n}{dx^2}+
\frac{1}{\rhotil(x)}\frac{d\si_n(x)}{dx}\frac{du_n}{dx}
\pr
\label{boltz14}
\eea 
The LHS of eq.(\ref{boltz11}) is written, using Navier-Stokes equation (\ref{boltz1}), as
\bea
\frac{1}{h}(u_n(x)-u_{n-1}(x))+A~u_{n-1}\frac{du_n}{dx}=        \nn
\frac{d}{dx}\left\{ \ka_{n-1}\frac{du_n}{dx} +A(u_n-u_{n-1})u_{n-1}  \right\} -m^2u_n-\frac{\la}{3!}{u_n}^3\com\q A=\frac{\sqrt{\la\si_0}}{|m|}
\pr
\label{boltz15}
\eea 
Generally we have some cases about the two quantities (\ref{boltz14}) and (\ref{boltz15}). \nl

Case W1:\ The two quantities coincide.\nl
In this case, the present theory is {\it well-defined}. There is no problem.\nl

Case W2:\ The 2 quantities differ, but they can be equal by changing the two constants, 
$m^2$ and $\la$. Namely,  
\bea
\ka_n(x)\frac{d^2u_n}{dx^2}+
\frac{1}{\rhotil_n(x)}\frac{d\si_n(x)}{dx}\frac{du_n}{dx} =  \nn
\frac{d}{dx}\left\{ \ka_{n-1}\frac{du_n}{dx} +A' (u_n-u_{n-1})u_{n-1}  \right\} 
-{m'}^2u_n-\frac{\la'}{3!}{u_n}^3\com\q A' =\frac{\sqrt{\la'\si_0}}{|m'|}
\pr
\label{boltz16}
\eea 
By identifying $m^2$ and $\la$ by $(m^2)_{n-1}$ and $\la_{n-1}$, and 
${m^2}'$ and $\la'$ by $(m^2)_n$ and $\la_n$, the above equation defines 
the {\it step-flow} of the two {\it constants} $(m^2)_n$ and $\la_n$. 
We can regard this flow as the {\it renormalization} flow by identifying 
the step-flow with {\it thermalization} to the equilibrium state. 
\footnote{
We regard the change $(m^2)_{n-1}\ra (m^2)_{n}, (\la)_{n-1}\ra (\la)_{n}$ 
as the renormalization along the step-flow ('time' development). The consistent 
model generally has the renormalization both in the space-distribution (\ref{fluct7}) and 
in the 'time' development. 
             }

Case W1 is the special case of W2, namely,  n('time')-independent one. 
We call the both cases {\it well-defined} theory. 
We call other cases, where eq.(\ref{boltz11}) is not valid, {\it ill-defined} theory.   
The validity check of eq.(\ref{boltz11}) selects the present model, (\ref{DTheat6b}), 
well-defined or not. 

Next we explain how eq.(\ref{boltz12}) is valid in the well-defined case. 
Noting the heat current $q_n(x)$ is, in the expression (\ref{boltz10}), the 
higher-moment of the distribution $f_n(x,v)$ than the pressure $P_n(x)$, we newly define 
the {\it heat conductivity} $\om_n(x)$ in the same way as (\ref{boltz13}). 
\bea
\mbox{[Comp 6]}\q \om_n(x)\ \mbox{Equation}\qqqq\qqqq  \nn
q_n(x)=\om_n(x)\frac{d\Tcal_n}{dx}
\pr
\label{boltz17}
\eea 
\footnote{
The heat conductivity $\om_n(x)$ appears here for the first time. 
We do not have this quantity in the starting energy expression eq.(\ref{DTheat6b}). 
In this point, the heat conductivity differs from the viscosity $\si_n(x)$.  
$\si$ is the basic physical quantity which characterize the present 
system, while $\om$ is the derived quantity from other basic ones.   
}
Noting the equation of state $P_n(x)=\kB\rho_n(x)\Tcal_n(x)$\ (\ref{boltz10}), 
eq.(\ref{boltz12}) is written as 
\bea
\half\{
\frac{1}{h}(P_n-\frac{\rho_n}{\rho_{n-1}}P_{n-1})+u_{n-1}\frac{dP_n}{dx}-
u_{n-1}\frac{d\rho_n}{dx}\frac{P_n}{\rho_n}
       \}
=-\frac{dq_n(x)}{dx}-P_n(x)\frac{d\un(x)}{dx}
\pr
\label{boltz18}
\eea 
Using the relations, $P_n=-\si_n\times du_n/dx$ and $q_n=\om_n\times d\Tcal_n/dx$, 
we obtain the differential equation for $\om_n(x)$ as
\bea
\mbox{[Comp 7]}\q \mbox{Eq.(\ref{boltz12})}\ \mbox{Check Equation}\qqqq\qqqq  \nn
\frac{d}{dx}(\om_n\frac{d\Tcal_n}{dx})=
\half\{
\frac{1}{h}(\si_n\frac{du_n}{dx}-\frac{\rho_n}{\rho_{n-1}}\si_{n-1}\frac{du_{n-1}}{dx})  \nn
+u_{n-1}\frac{d}{dx}(\si_n\frac{du_n}{dx})-
u_{n-1}\frac{d\rho_n}{dx}\frac{\si_n}{\rho_n}\frac{du_n}{dx}
       \}                                  \q
+\si_n(\frac{d\un(x)}{dx})^2
\pr
\label{boltz19}
\eea 
This differential equation determines $\om_n(x)$ with no use of $q_n(x)$. 
Hence eq.(\ref{boltz19}) fixes $\om_n(x)$ independently of eq.(\ref{boltz17}). 
The (numerical) equality check of $\om_n(x)$ obtained from both equations 
is the validity check of eq.(\ref{boltz12}). When the equality do hold or do not, 
we call the present model, (\ref{DTheat6b}), {\it well-defined} or {\it ill-defined} respectively.

In this section, we presents how to calculate 
$\rho_n$ (\ref{boltz4b}), $P_n=\kB\rho_n\Tcal_n$ (\ref{boltz11}), $q_n$ (\ref{boltz12}), $\si_n$ (\ref{boltz13}), $\om_n$ (\ref{boltz17})  with no use of the distribution function $f_n(x,v)$. 
We can do the numerical simulation of them.  

In the remaining sections, we present an alternative approach to the distribution function $f_n(x,v)$.

\begin{figure}
\caption{
The harmonic oscillator with friction, (\ref{qm3}). 
        }
\begin{center}
\includegraphics[height=6cm]{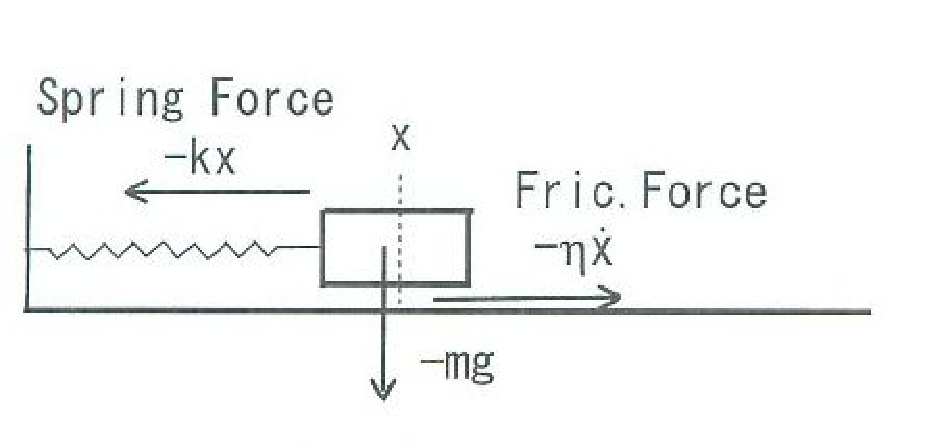}
\end{center}
\label{HOmodel}
\end{figure}
\section{Classical and Quantum Mechanics and Its Trajectory Geometry\label{qm}}

We can treat the classical mechanics and 
its quantization ( the quantum mechanics, not the quantum field theory) in the same way. 
In this case, 
the model is simpler than the previous case (space-field theory) and we can see 
the {\it geometrical structure} clearly. 
Let us begin with the energy function of a system variable , $x$, 
(1 degree of freedom). For example the {\it position} (in 1 dimensional space) 
of the harmonic oscillator with friction. 
We take the following $n$-th energy function to define the step flow. 
\bea
K_n(x)=  V(x)
+\frac{\eta}{2h}(x-x_{n-1})^2
+\frac{m}{2h^2}(x-2x_{n-1}+x_{n-2})^2+K_n^0
\com
\label{qm1}
\eea 
where $V(x)$ is the general potential and $K_n^0$ is a constant which does not depend on $x$. 
For the harmonic oscillator $V(x)=kx^2/2$ where $k$ is the spring constant. 
$\eta$ is the viscosity and $m$ is the particle mass. 
\footnote{
Here we list the dimension of parameters and variables in this section. 
[$x$]=[$x_n$]=L, [$v$]=[$v_n$]=L/T, [$t$]=[$t_n$]=T, [$q$]=T$^{1/2}$M$^{-1/2}$, [$\hbar$]=ML$^2$/T, 
[$m$]=M, [$\eta$]=M/T, [$h$]=T, [$K_n$]=[$V$]=ML$^2$/T$^2$, [$k$]=M/T$^2$, [$\sqrt{\eta h}$]=M$^{1/2}$, 
[$\sqrt{mh^2}$]=M$^{1/2}$T. Some ones, such as $t, m, h, q$ and $V$, appearing before this section 
have different dimensions in this section. 
}
We assume $x_{n-1}$ and $x_{n-2}$ are given values. As in Sec.\ref{DTheat}, the n-th step position $\xn$ is 
given by the {\it minimal principle} of the n-th energy function $K_n(x)$:\ $\del K_n=0, x\ra x+\del x$. 
\bea
\left.\frac{\del V}{\del x}\right|_{x=\xn}+\frac{\eta}{h}(\xn-x_{n-1})+\frac{m}{h^2}(\xn-2x_{n-1}+x_{n-2})=0
\com
\label{qm2}
\eea 
This is the recursion relation among three quantities, $x_n$, $x_{n-1}$ and $x_{n-2}$. 
\footnote{
Simulation results of this equation is later given. 
}
With the time $t_n$ (\ref{DTheat4}), the continuous limit ($h\ra 0$) gives us
\bea
\frac{dV(x)}{dx}+\eta\frac{dx}{dt}+m\frac{d^2x}{dt^2}=0
\com
\label{qm3}
\eea 
where $t_n=nh\ra t,\ \xn=x(t_n)\ra x(t),\ 
(\xn-x_{n-1})/h =dx/dt|_{t_n}\ra dx/dt,\ (\xn-2x_{n-1}+x_{n-2})/h^2 =d^2x/dt^2|_{t_n}\ra d^2x/dt^2\ $. For the case of $V=kx^2/2$, this is the harmonic oscillator with the friction \ $\eta$. 
See Fig.\ref{HOmodel}. This is 
a simple {\it dissipative} system. 
\footnote{
The eq. (\ref{qm3}) is compared with the eq. (\ref{DTheat6}) for the case of 
the no external force (and $\si=\si_0$(const),$\rhotil=\rhotil_0$(const)):\ 
$$
\rhotil_0\frac{\del V(u(x, t))}{\del u(x, t)}-\si_0 \frac{\pl^2 u(x,t)}{\pl x^2}+\rhotil_0\frac{\pl u(x,t)}{\pl t}=0
\com
$$
where we notice the friction term in eq.(\ref{qm3}) corresponds to the dissipative term in eq.(\ref{DTheat6}). 
}

  The recursion relation (\ref{qm2}) gives us, for the initial data $x_0$ and $x_1$, 
the series \{$\xn =x(t_n)~ |n=0,1,2,\cdots$ \}. This is the classical 'path'. The fluctuation 
of the path comes from the {\it uncertainty principle} of the {\it quantum mechanics} in this case. 
(We are treating the system of 1 degree of freedom. No statistical procedure is necessary. ) 
In the quantum effect, the energy uncertainty grows as $\Del t\cdot \Del E\geq \hbar$. 
Hence the path $\xn$, obtained by the recursion relation (\ref{qm2}), has 
quantum uncertainty. 
As in Sec.\ref{fluct}, we can generalize the n-th energy function $K_n(x)$, (\ref{qm1}), to 
the following one $\Ga(x_{n-1}, x_{n-2})$ in order to take into account the quantum effect. 
\bea
\e^{-\frac{1}{\hbar}h \Ga(x_{n-1}, x_{n-2})}=\int_{-\infty}^{\infty} dx~\e^{-\frac{1}{\hbar}h K_n(x)}\com\nn
K_n(x)=  V(x)
+\frac{\eta}{2h}(x-x_{n-1})^2
+\frac{m}{2h^2}(x-2x_{n-1}+x_{n-2})^2
+K_n^0
\pr
\label{qm4}
\eea 
We can evaluate the {\it quantum} effect by the expansion around 
the classical value $\xn$\ :\ $x=\xn +\sqrt{\hbar}~ q$ where $\hbar$ is Planck constant. 
\footnote{
Do not confuse $\hbar$ (Planck constant/2$\pi$) with $h$ (1 step time-interval). 
}
\bea
\e^{-\frac{1}{\hbar}h\Ga(\xn; x_{n-1}, x_{n-2})}=\int dx~\e^{-\frac{1}{\hbar}h K_n(x)}
=\int dq~\e^{-\frac{1}{\hbar}h K_n(\xn+\sqrt{\hbar} q)}\com\nn
\left.\frac{\del K_n}{\del x}\right|_{x=\xn}=
\left.\frac{\del V}{\del x}\right|_{x=\xn}+\frac{\eta}{h}(\xn-x_{n-1})+\frac{m}{h^2}(\xn-2x_{n-1}+x_{n-2})=0\com\nn
\Ga_n\equiv\Ga(\xn; x_{n-1}, x_{n-2})=K_n(\xn)+\frac{\hbar}{2h}\ln(k+\frac{\eta}{h}+\frac{m}{h^2})
\com
\label{qm5}
\eea 
where the final expression is for the oscillator model: $V=kx^2/2$. The quantum effect 
does not depend on the step number $n$. It means the quantum effect contributes to the energy 
as an additional fixed constant at each step. 

The energy rate is obtained as
\bea
h \frac{dK(\tn)}{d\tn}\equiv K_{n+1}(x_{n+1})- K_{n}(x_{n})=\Ga_{n+1}-\Ga_{n}
\equiv h \frac{d\Ga(\tn)}{d\tn}\nn
=V(x_{n+1})-V(\xn)+
\frac{\eta}{2h}\{ (x_{n+1}-x_{n})^2 - (x_{n}-x_{n-1})^2 \}  \nn
+\frac{m}{2h^2}\{  (x_{n+1}-2x_{n}+x_{n-1})^2  - (x_{n}-2x_{n-1}+x_{n-2})^2 \} 
K_{n+1}^0-K_n^0
\pr
\label{qm5b}
\eea 
The present system is again an open system, and the energy generally changes. 

In terms of the position difference $\xn-x_{n-1}\equiv \Del x_n$ and 
the velocity difference $(\xn-2x_{n-1}+x_{n-2})/h\equiv v_n-v_{n-1}\equiv \Del v_n$, 
we can rewrite the energy at $n$-step and 
read the {\it metric} as follows. 
\footnote{
 $v_n\equiv (\xn-x_{n-1})/h \ = \Del x_n/h$
}
\bea
K_n(\xn)=  V(\xn)
+\frac{\eta}{2h}(\xn-x_{n-1})^2
+\frac{m}{2h^2}(\xn-2x_{n-1}+x_{n-2})^2+K_n^0\nn
=\frac{1}{h^2}\{
V(\xn)(\Del t)^2
+\frac{\eta h}{2}(\Del \xn)^2
+\frac{m h^2}{2}(\Del v_n)^2
                   \}+K_n^0
\com
\label{qm6}
\eea 
where $h$ (time increment) in the first term within the round brackets is replaced by $\Del t$. 
This shows us the metric for the n-step energy function is given by
\bea
(\Del s_n)^2\equiv 2h^2(K_n(\xn)-K_n^0)
=
2V(\xn'/\sqrt{\eta h})(\Del t)^2
+(\Del \xn')^2
+(\Del v_n')^2
                                      \com\nn 
\xn'\equiv \sqrt{\eta h}\xn\com\q {v_n}'\equiv \sqrt{mh^2}v_n
\com
\label{qm7}
\eea 
where , for the oscillator model, $V(\xn'/\sqrt{\eta h})=(k'/2){\xn'}^2, k'\equiv k/\eta h$. 
Eq.(\ref{qm7}) shows the energy-line element ${\Del s}^2$ in the ($t, \xn', {v_n}'$) space. 
\footnote{
In eq.(\ref{qm7}), the first term shows the potential part, the second one the kinetic part 
and the third one a new term. In ref.\cite{ICSF2010} and ref.\cite{SI1004}, the hysteresis term appears 
as a new one. 
}
Note that the above metric is {\it along the path}  
$x_n=x(t_n),\ v_n=v(t_n)=(x(t_n)-x(t_{n-1}))/h$ given by (\ref{qm2}). 
The metric is used, in the next section, as the {\it geometrical} basis for 
fixing the statistical ensemble.   

We define the system energy \ul{SysE}$_n$, the dissipative energy \ul{DisE}$_n$ 
and the constant $K_n^0$ as follows. 
\bea
\mbox{\ul{SysE}}_n =  \frac{1}{m}V(\xn)+\frac{1}{2h^2}(\xn-2x_{n-1}+x_{n-2})^2\com\q
\mbox{\ul{DisE}}_n =  \frac{\eta}{2hm}(\xn-x_{n-1})^2\com\nn
K_n^0= 0     
\com\q
K_n(\xn)=m\mbox{\ul{SysE}}_n + m\mbox{\ul{DisE}}_n 
\pr
\label{qm8}
\eea 
\ul{DisE}$_n$ expresses the hysteresis (non-Markovian) energy at the n-th step. 
We can obtain the dynamical energy 
(the potential energy plus the kinetic energy)  \ul{DynE}$_n$ as
\bea
{ \mbox {\ul{DynE}} }_n= \frac{1}{m}V(\xn)+\frac{1}{2h^2}(\xn-x_{n-1})^2
\pr
\label{qm9}
\eea 

\begin{figure}
\caption{
The dissipative energy \ul{DisE}$_n$, (\ref{qm8}), of the frictional harmonic oscillator Fig.\ref{HOmodel} with the parameters (\ref{qm10}) . 
        }
\begin{center}
\includegraphics[height=6cm]{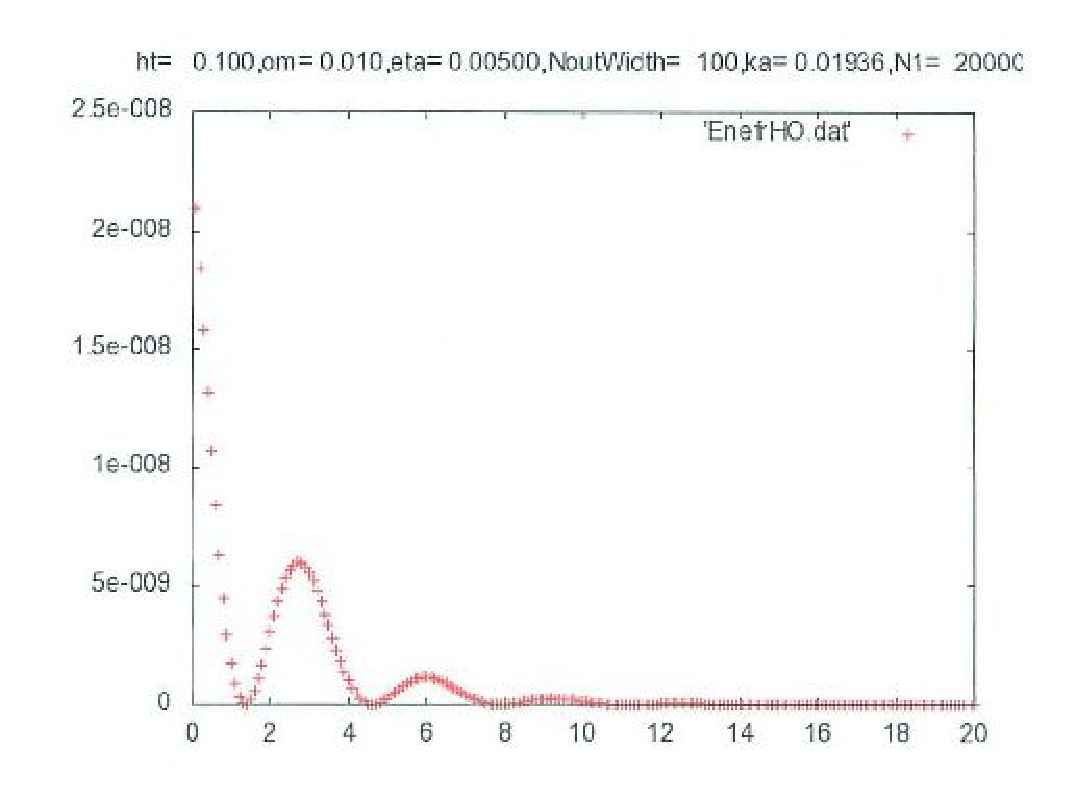}
\end{center}
\label{EnefrHO}
\end{figure}

\begin{figure}
\caption{
The system energy \ul{SysE}$_n$, (\ref{qm8}), of the frictional harmonic oscillator Fig.\ref{HOmodel} with the parameters (\ref{qm10}) . 
        }
\begin{center}
\includegraphics[height=6cm]{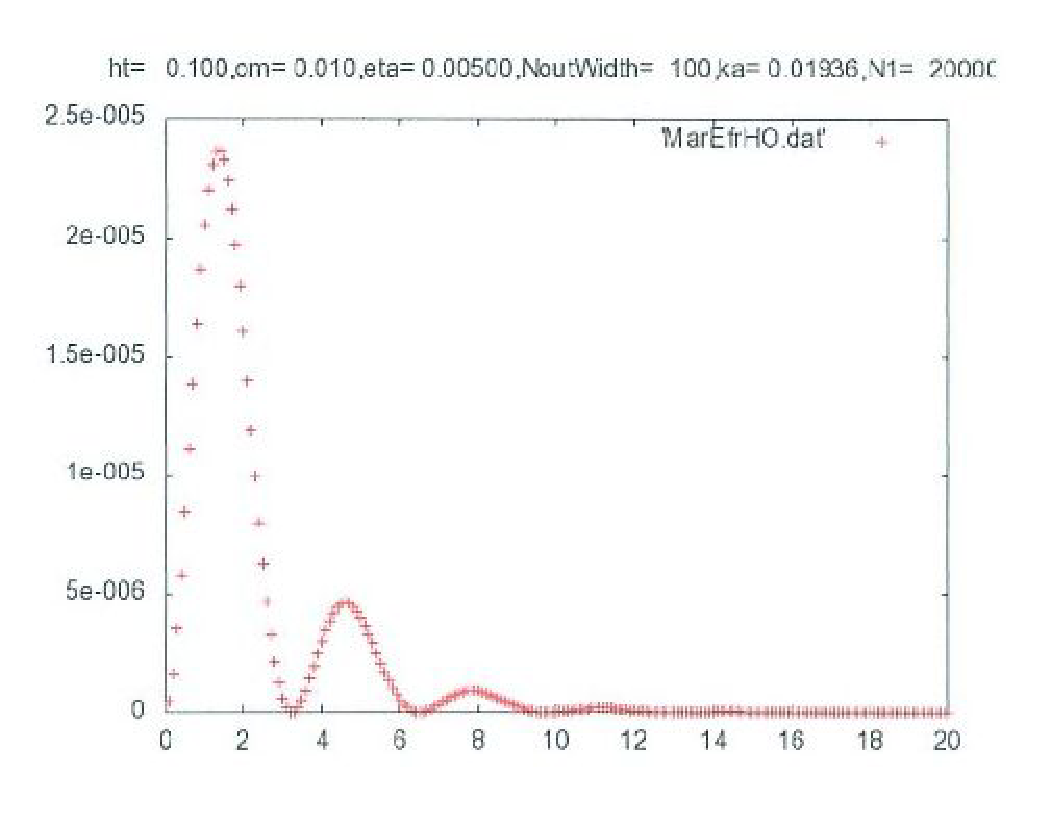}
\end{center}
\label{MarEfrHO}
\end{figure}

\begin{figure}
\caption{
The dynamical energy \ul{DynE}$_n$, (\ref{qm9}), of the frictional harmonic oscillator Fig.\ref{HOmodel} with the parameters (\ref{qm10}) . 
        }
\begin{center}
\includegraphics[height=6cm]{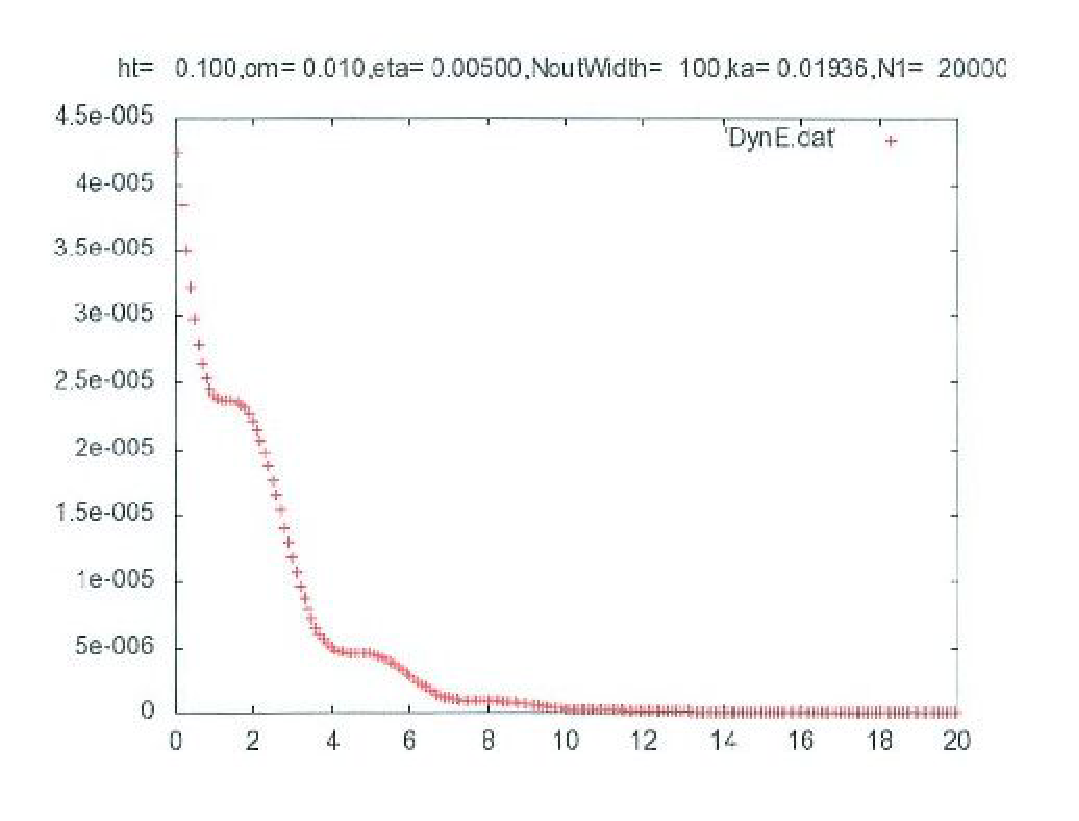}
\end{center}
\label{DynE}
\end{figure}

For the special case of the following, we list the simulation results here and in App.C. 
The horizontal axis is $t_n\om=nh\om$. 
\bea
h=0.1\com\q\om=\sqrt{k/m}=0.01\com\q \eta'=\eta/m=0.005 (\mbox{Elasticity dominate})\com\nn
 0\leq t_n\om=nh\om \leq 20\com\q  \mbox{total step no} =20000. 
\pr
\label{qm10}
\eea 
The movement $x_n$ and the velocity distribution $v_n=(x_n-x_{n-1})/h$ are shown 
in App.C.  
The dissipative energy \ul{DisE}$_n$  and the system energy 
\ul{SysE}$_n$ are shown in Fig.\ref{EnefrHO} and Fig.\ref{MarEfrHO} respectively. They are 
, in the oscillatory way, damping. The dynamical energy \ul{DynE}$_n$ is shown in Fig.\ref{DynE}.
It damps, not in the oscillatory way but in the stick-slip way. 

An advantageous point of the step-wise solution (\ref{qm2}) over 
the analytic one of (\ref{qm3}), is that we need {\it not} treat the 3 cases, 
$4k/m~>~\eta^2/m^2$ (elasticity dominate),\ 
$4k/m~<~\eta^2/m^2$ (viscosity dominate) and 
$4k/m~=~\eta^2/m^2$ (resonant), 
separately. 
This is because (\ref{qm2}) is linear with repect to (w.r.t.) $x_n$, whereas (\ref{qm3}) is the 
second-order equation w.r.t. $d/dt$. 

\section{Statistical Ensemble, Geometry and Initial Condition\label{se} }
In this section, we consider a statistical ensemble of the classical mechanical 
system taken in the previous section. Namely, we take $N$ 'copies' of the classical 
model and regard them as a set of  (1 dimensional) particles, where 
the dynamical configuration distributes in the probabilistic way. 
$N$ is a large number. 
\footnote{
For example, $N\sim 10^{23}$. 
We are modeling the many-body ($N$: large) system by the statistical collection (ensemble) 
of many copies of one-body system (\ref{qm2}) or (\ref{qm3}). 
}
The set has $N$ degrees of freedom:\ $x_1, x_2, \cdots, x_N$. 
As the physical systems, (1 dimensional) {\it viscous} gas and  {\it viscous} liquid are  examples. 
\footnote{
We are considering $N$-body problem where each particle moves (fluid flows) with moderate friction. 
We aproach it using the effective 1-body energy function (\ref{qm1}). 
} 
Each particle obeys the (step-wise) Newton's law (\ref{qm2}) with different 
{\it initial conditions}. 
$N$ is so large that we do not or can not observe the initial data. Usually we do not have interest 
in the trajectory of every particle and do not observe it. We have interest only in the macroscopic 
quantities such as {\it total energy} and {\it total entropy}. The N particles (fluid molecules) in the
present system are "moderately" interacting each other in such a way that each particle almost independently  moves except that energy is exchanged.

\begin{figure}
\caption{
The path $\{(y(t),w(t),t) | 0\leq t\leq \be  \}$ of line in 3D bulk space (X,P,t). 
        }
\begin{center}
\includegraphics[height=8cm]{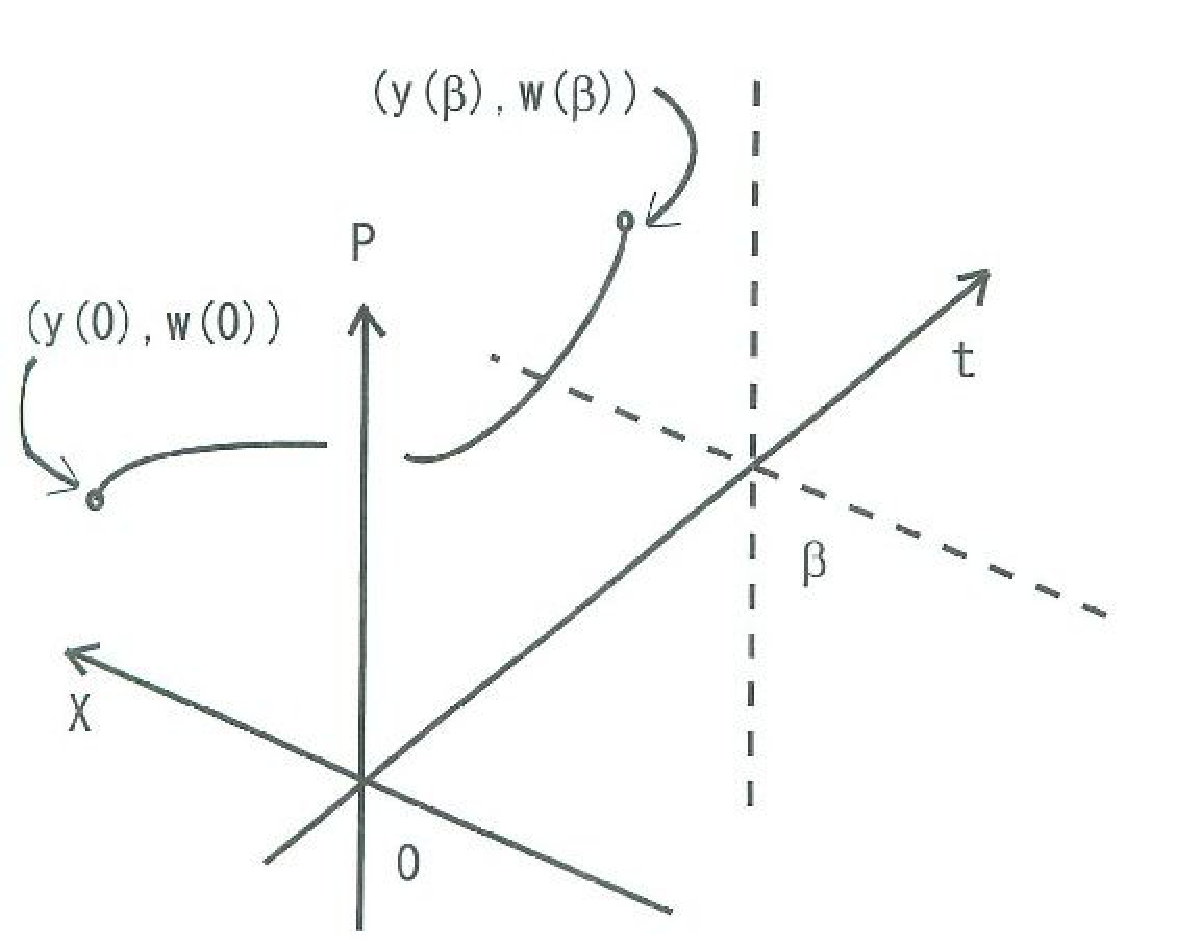}
\end{center}
\label{LinePath}
\end{figure}
As the statistical ensemble, we adopt the Feynman's idea of "path-integral"
\cite{SI0801, SI0812, SI0903Singa, SI0909, SI0912,SI1205,SI1004,ICSF2010}. 
We take into account all possible paths $\{ y_n \}$.  $\{ y_n \}$ need not satisfy (\ref{qm2}) nor 
certain initial condition. 
As the {\it measure} for the summation (integral) over all paths, 
we propose the following ones based on the geometry of (\ref{qm7}).  
As the first measure, we construct it in terms of the {\it length}, using the "Dirac-type" metric
\cite{SI1004,ICSF2010}. 
\bea
(ds^2)_D\equiv 2V(X)dt^2+dX^2+dP^2\q -\mbox{on-path}(X=y(t), P=w(t))\ra \nn 
(2V(y)+\ydot^2+\wdot^2) dt^2\com\nn
L_D=\int_0^\be ds|_{on-path}=\int_0^\be\sqrt{2V(y)+\ydot^2+\wdot^2}dt=
h \sum_{n=0}^{\be/h}\sqrt{2V(y_n)+\ydot_n^2+\wdot_n^2}\com \nn
d\mu=\e^{-\alinv L_D}\Dcal y\Dcal w\com\q
\e^{-\be F}=\int\prod_n dy_n dw_n\e^{-\alinv L_D}
\com
\label{se1}
\eea 
where $\al$ is a parameter with the dimension of length ([$\al$]=L) and $V(X)=(k/2)X^2$. 
See Fig.\ref{LinePath}. 
As explained in Sec.\ref{fluct}, it is appropriately chosen problem by problem. 
$\be$ is introduced to restrict the $t$-region ($0\leq t\leq \be$) and, in this context, 
should be regarded as a part of the choice of the ensemble. $\be$ plays the role 
of the inverse temperature. 
\footnote{
$\be/h=N$ should be an (large) integer. The increment $h$ is the (inverse) temperature unit as well as the time unit. 
From the dimensional analysis ${k_B}^{-1}\eta \ltil^2/\be$ corresponds to the temperature. 
Here $k_B$ is Boltzmann's constant and $\eta \ltil^2\equiv \hbar'$ is the combination 
of the friction coefficient $\eta$ and some length scale $\ltil$ ([$\ltil$]=L) such as the mean free path 
of the fluid particle. 
Note that $\hbar'$ 
has the same dimension as $\hbar$. [$\hbar'$]=[$\hbar$]=ML$^2$/T. 
}
Among all possible paths $\{ y_n\}$, the minimal length ($\del L_D=0$) solution, (\ref{qm2}), 
gives the dominant path $\{ x_n\}$. 

The second choice is constructed using the "Standard-type" metric.
\bea
(ds^2)_S\equiv \frac{1}{dt^2}[(ds^2)_D]^2\q -\mbox{on-path}(X=y(t), P=w(t))\ra \nn 
(2V(y)+\ydot^2+\wdot^2)^2 dt^2\com\nn
L_S=\int_0^\be ds|_{on-path}=\int_0^\be (2V(y)+\ydot^2+\wdot^2)dt=
h \sum_{n=0}^{\be/h}(2V(y_n)+\ydot_n^2+\wdot_n^2)\com\nn
d\mu=\e^{-\alinv L_S}\Dcal y\Dcal w\ ,\ 
\e^{-\be F}=\int\prod_n dy_n dw_n\e^{-\alinv L_S}
=(\mbox{const})\int\prod_{n} dy_n\e^{-\frac{h}{\al}(2V(y_n)+\ydot_n^2)}\ ,
\label{se2}
\eea 
where we should notice $dt$ ($=h> 0$) is non-zero. In both cases above we take 
the metric of the 3 dimensional (bulk) space-time (X, P, t) , which is consistently 
chosen with the trajectory metric (\ref{qm7}). 
Note that the standard case has the same expression as the free energy 
(trace of the {\it density matrix})  expression in the Feynman's textbook\cite{Fey72}.

\begin{figure}
\caption{
The two dimensional surface, (\ref{se3}),  in 3D bulk space (X,P,t). 
        }
\begin{center}
\includegraphics[height=8cm]{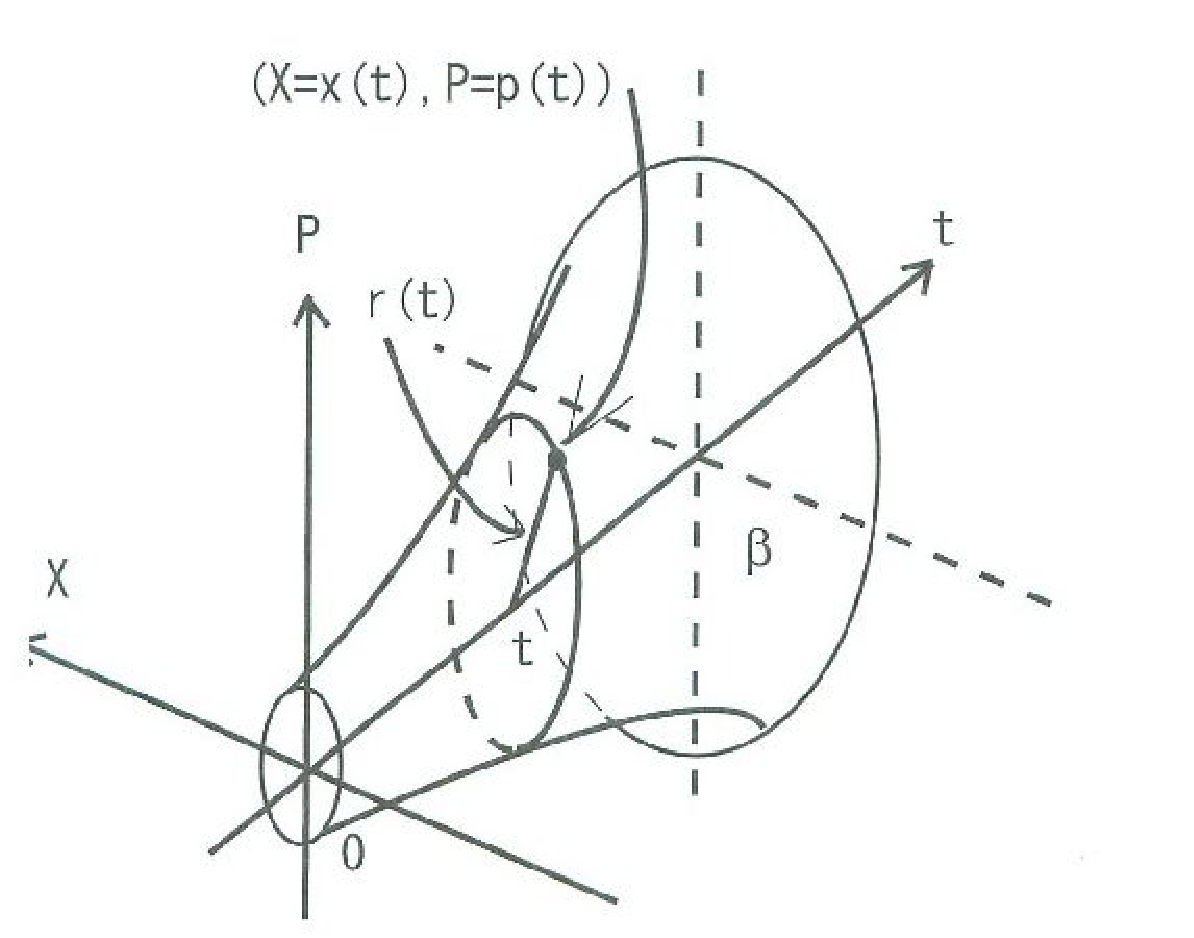}
\end{center}
\label{2DHyperSurf}
\end{figure}
Another choice of path is making use of {\it surfaces} instead of {\it lines}. 
Let us consider the following 2 dim surface in the 3 dim manifold ($X, P, t$). 
Assuming Z$_2$ invariance both in $X$ and in $P$, the general one is 
\bea
\frac{X^2}{a(t)^2}+\frac{P^2}{b(t)^2}=1\com\q 0\leq t\leq \be
\com
\label{se3}
\eea 
where $a(t)$ and $b(t)$ are arbitrary (positive) functions of $t$. 
We take $a(t)=b(t)\equiv r(t)$ for simplicity. 
See Fig.\ref{2DHyperSurf}. 
By varying the form of $\{ r(t):\ 0\leq t\leq \be\}$, we obtain different surfaces. 
Regarding each of them as a path used in the Feynman's path-integral, we obtain 
the following statistical ensemble. First the {\it induced metric} $g_{ij}$ on the surface (\ref{se3}) 
is given as 
\bea
\left.(ds^2)_D\right|_{\mbox{on-path}}=\left. 2V(X)dt^2+dX^2+dP^2\right|_{\mbox{on-path}}
=\sum_{i,j=1}^{2}g_{ij}dX^idX^j\com\nn
(g_{ij})=\left(
\begin{array}{cc}
1+\frac{2V}{r^2\rdot^2}X^2& \frac{2V}{r^2\rdot^2}X P \\
\frac{2V}{r^2\rdot^2}P X & 1+\frac{2V}{r^2\rdot^2}P^2
\end{array}
          \right)
\com
\label{se4}
\eea 
where  $(X^1, X^2)=(X, P)$. Then the {\it area} of the surface (\ref{se3}) is given by
\bea
A=\int\sqrt{\det g_{ij}}d^2X=\int\sqrt{1+\frac{2V}{\rdot^2}}dX dP
\pr
\label{se5}
\eea 
We consider all possible surfaces of (\ref{se3}). The statistical distribution 
is, using the {\it area} $A$, given by 
\bea
\e^{-\be F}=\int_0^\infty d\rho\int_{\begin{array}{c}
                                                        r(0)=\rho \\
                                                        r(\be)=\rho
                                                 \end{array}}
\prod_t\Dcal X(t)\Dcal P(t)\e^{-\frac{1}{\al}A}
\pr
\label{se6}
\eea 

In relation to Boltzmann's equation (Sec.\ref{boltz}), we have directly defined 
the distribution function $f(t,x,v)$ using the geometrical quantities in the 
3 dim bulk space. Three statistical ensembles are proposed. 
In order to select which one is the most appropriate, it is necessary to 
{\it numerically} evaluate the models with the proposed ensembles  
and compare the result with the real data 
appearing both in the natural phenomena and in the laboratory experiment.

In App.D, another model called "Spring-Block" model is explained. This is 
the simplified model of the earthquake. The same thing in this section is valid 
by taking the potential $V(X)$ as 
\bea
V(X)=\frac{k'}{2}X^2+(\lbar'-\Vbar't)k'X\ ,\ k'=\frac{k}{\eta h}\ ,\ 
\lbar'=\sqrt{\eta h}\lbar\ ,\ \Vbar'=\sqrt{\eta h}\Vbar
\ .
\label{se7}
\eea 

\section{Conclusion\label{conc} }

We have presented the field theory approach to Boltzmann's transport 
equation. 
The collision term is explicitly obtained. 
Time is {\it not} used, instead the step number $n$ plays the role. 
We have presented the $n$-th energy functional (\ref{DTheat6b}) which 
gives the step $n$ configuration $\un(x)$  
from the minimal energy principle. 
We regard the step flow ( the increase of $n$ ) as the evolution of 
the system, namely, {\it time-development}. 
Burgers (Navier-Stokes) equation is obtained by identifying time $t$ 
as $nh$ (\ref{DTheat4}).  Fluctuation effect due to the micro structure and 
micro (step-wise) movement is taken into account by generalizing the $n$-th energy functional 
$\Itil_n[u(x)]$, (\ref{DTheat6b}), to $\Ga[u(x); u_{n-1}(x)]$, (\ref{fluct1}),  
where the classical path $\un(x)$ is dominant but all possible paths 
are taken into account (path-integral). 
{\it Renormalization}, due to the statistical fluctuation, is explicitly done. The total 
energy generally does {\it not} conserve. The system is an open one, namely, the energy 
comes in from or go out to the outside. 
In the latter part of the text, we have presented a direct approach to the distribution function $f_n(x,v)$ 
based on the {\it geometry} emerging from the mechanical (particle-orbit) dynamics. 
We have examined the {\it dissipative} 
system using the {\it minimal (variational) principle} which is the key principle in the standard field theory\cite{YM526879}.


\section{Appendix A\ \ (3+1)D Scalar Field Theory\label{sf} }

3+1 dimensional scalar field is here treated in the present step-wise formalism. 
We start with the following $n$-th step energy functional.
\bea
I_n[\phi(\bx)]=\intthx \{ \frac{1}{2}( {\bna} \phi)^2 +V(\phi)
+\frac{1}{2h^2}(\phi-2\phi_{n-1}+\phi_{n-2})^2\} \q n=2, 3, \cdots\com \nn
V(\phi)=\frac{m^2}{2}\phi^2+\frac{\la}{4!}\phi^4\com\q \phi=\phi(\bx)\ ,\ \phi_{n-1}=\phi_{n-1}(\bx) ,\ 
\phi_{n-2}=\phi_{n-2}(\bx)
\com
\label{sf1}
\eea 
where $\phi_{n-2}$ and $\phi_{n-1}$ are given.  $(\bx)=(x_1, x_2, x_3)$ is the 3 dimensional spacial 
coordinates. The step $n$ configuration $\phin(\bx)$ is defined to be the energy minimal one.
\bea
\left.\frac{\del I_n}{\del\phi}\right|_{\phi=\phin}=
-{\bna}^2\phin +\left.\frac{\del V}{\del \phi}\right|_{\phi=\phin}
+\frac{1}{h^2}(\phin-2\phi_{n-1}+\phi_{n-2})=0 \com \nn
\frac{\del V}{\del \phi}=m^2\phi+\frac{\la}{3!}\phi^3
\pr
\label{sf2}
\eea 
Using the step-time notation:\ $\phin(\bx)\equiv \phi(\bx, \tn),\ \tn=nh$, 
we obtain, in the continuous time limit $h\ra +0$, 
\bea
(\pl_t^2-\bna^2+m^2)\phi+\frac{\la}{3!}\phi^3=0
\pr
\label{sf3}
\eea 
This is the (3+1) dim $\phi^4$ scalar field equation.


\section{Appendix B\ \ Calculation of Fluctuation Effect\label{fluctcal} }

In Sec.\ref{fluct}, we have developed the method of calculating the statistical 
fluctuation occurring in the (1-dim) viscous fluid matter. 
The background-field method is taken. At the 1-loop approximation, 
the key quantity to calculate the energy functional $\Ga[u(x)]$ is 
the {\it heat-kernel} $G(x,y;\tau)$ given by (see eq.(\ref{fluct5}))
\bea
<x|\e^{-\tau D}|y>\equiv G(x,y;\tau)\com\nn
(\frac{\pl}{\pl\tau}+D)G(x,y;\tau)=0\com\q \lim_{\tau\ra +0}G(x,y;\tau)=\del(x-y)\com\nn
D=-\ep^{-1}\frac{\pl^2}{\pl x^2} + 2u_{n-1}(x) A\frac{d}{dx} -\Vbar(x)\com\q 
\Vbar(x)=-\la {u_n(x)}^2-m^2-\frac{1}{h}
                      \com
\label{fluctcal1}
\eea 
where $|x>$ and $<x|$ are Dirac's abstract state vectors. 
$\ep\equiv\rhotil_0/\si$ is explicitly written to show the dimension consistency. 
We take $A=0$ in this appendix. 
In the text, we take $\ep=1$. For the calculation, in this appendix, 
we change the scale of $\tau$ and $D$ as follows.
\bea
\tau\q\ra\q \frac{\tau}{\ep}=\tautil\com\q D\q\ra\q \ep D=\Dtil\com\nn
\Dtil=-\frac{\pl^2}{\pl x^2}-\ep \Vbar(x)\com\q \mbox{[$\tautil$]=L$^2$}\com\q
\mbox{[$\Dtil$]=L$^{-2}$}\com\q \mbox{[$\ep$]=1/LM}
                      \pr
\label{fluctcal1b}
\eea 
In the following within this appendix, for simplicity we omit the symbol of 'tilde'.  

The kernel is formally 
solved as
\bea
G(x,y;\tau)=G_0(x-y;\tau)+\int dz\int_0^\tau d\om S(x-z;\tau-\om)\ep\Vbar(z)G(z,y;\om)
\com
\label{fluctcal2}
\eea 
where $G_0(x-y;\tau)$ and the (heat-)propagator $S(x-y;\tau)$ are given by
\bea
G_0(x-y;\tau)=\int_{-\infty}^\infty\frac{dk}{2\pi}\e^{-k^2\tau+ik(x-y)}=
\frac{1}{\sqrt{4\pi\tau}}\e^{-\frac{(x-y)^2}{4\tau}}\com\q \tau>0\com\nn
S(x-y;\tau)=\int\int_{-\infty}^\infty\frac{dk}{2\pi}\frac{dk^0}{2\pi}
\frac{  \exp\{-ik^0\tau+ik(x-y) \}  }{-ik^0+k^2}
=\sh(\tau)G_0(x-y;\tau)
\pr
\label{fluctcal3}
\eea 
They satisfy
\bea
(\frac{\pl}{\pl\tau}-\frac{\pl^2}{\pl x^2})G_0(x-y;\tau)=0\com\q \tau>0\com\q
\lim_{\tau\ra +0}G_0(x-y;\tau)=\del (x-y)\com\nn
(\frac{\pl}{\pl\tau}-\frac{\pl^2}{\pl x^2})S(x-y;\tau)=\del(\tau)\del(x-y)\com\q
\lim_{\tau\ra +0}S(x-y;\tau)=\del (x-y)
\pr
\label{fluctcal4}
\eea 
Up to the first order of $\Vbar$, 
\bea
G(x,x;\tau)=\frac{1}{\sqrt{4\pi\tau}}+\int dz\int_0^\tau d\om S(x-z;\tau-\om)\ep\Vbar(z)
G_0(z-x;\om)+O(\Vbar^2)
\pr
\label{fluctcal5}
\eea 
The second term is evaluated as
\bea
\int dz\int_0^\tau d\om G_0(x-z;\tau-\om)\ep\Vbar(z)G_0(z-x;\om)=               \nn
\int dz\int_0^\tau d\om \frac{1}{4\pi}\frac{1}{\sqrt{(\tau-\om)\om}}\ep\Vbar(z)
\exp\{ -\frac{\tau}{4(\tau-\om)\om}(x-z)^2 \}
\pr
\label{fluctcal6}
\eea 
Finally the contribution to 
$\ln(\det D)^{-1/2}=\half\int_0^\infty(d\tau/\tau)\Tr G(x,y;\tau)=$\nl 
$\half\int_0^\infty(d\tau/\tau)\int_{-l}^l dx G(x,x;\tau)$ is evaluated as
\bea
\half\int_0^\infty d\tau\half\frac{1}{\sqrt{\tau\pi}}\int dz\ep\Vbar(z)=
\frac{1}{4\sqrt{\pi}}\int_0^{\ep^{-1}\mu^{-1}}\tau^{-1/2}d\tau\int dz\ep\Vbar(z)=\nn
\frac{l}{\sqrt{\pi}}\sqrt{\ep\La}-\frac{1}{2\sqrt{\pi\ep\mu}}\int_{-l}^l dz 
\ep\left( \la\un(z)^2+m^2+\frac{1}{h} \right)
\com
\label{fluctcal7}
\eea 
where the infrared cut-off parameter $\mu\equiv \sqrt{\si}/l$ and  
the ultraviolet cut-off parameter $\La\equiv h^{-1}$ are introduced. 
\footnote{
The dimensions of these parameters are [$\mu$]=[$\La$]=M/L. 
The space-integral part ($\int dx\cdots$) in (\ref{fluctcal7}) is evaluated as 
$\int_{-l}^l dx \exp \{ -(\tau/4(\tau-\om)\om)(x-z)^2 \} \sim \int_{-\infty}^\infty dx \exp\{ '' \}
=2\sqrt{\pi(\tau-\om)\om/\tau}$ where $l$ is safely extended to infinity.  
}

\section{Appendix C\ \ Simulation of Frictional Harmonic Oscillator  \label{Siml} }

In Sec.\ref{qm}, we take the frictional harmonic model Fig.\ref{HOmodel}. 
Some simulation results (Fig.\ref{EnefrHO}, Fig.\ref{MarEfrHO}, Fig.\ref{DynE}) are 
shown there. In this appendix, we show additional results. 

\begin{figure}
\caption{
The movement $x_n$, (\ref{qm2}), of the frictional harmonic oscillator Fig.\ref{HOmodel} with the parameters (\ref{qm10}) (Elasticity dominate). 
The step-wise solution reproduces the analytic solution (\ref{Sim1}).  
        }
\begin{center}
\includegraphics[height=6cm]{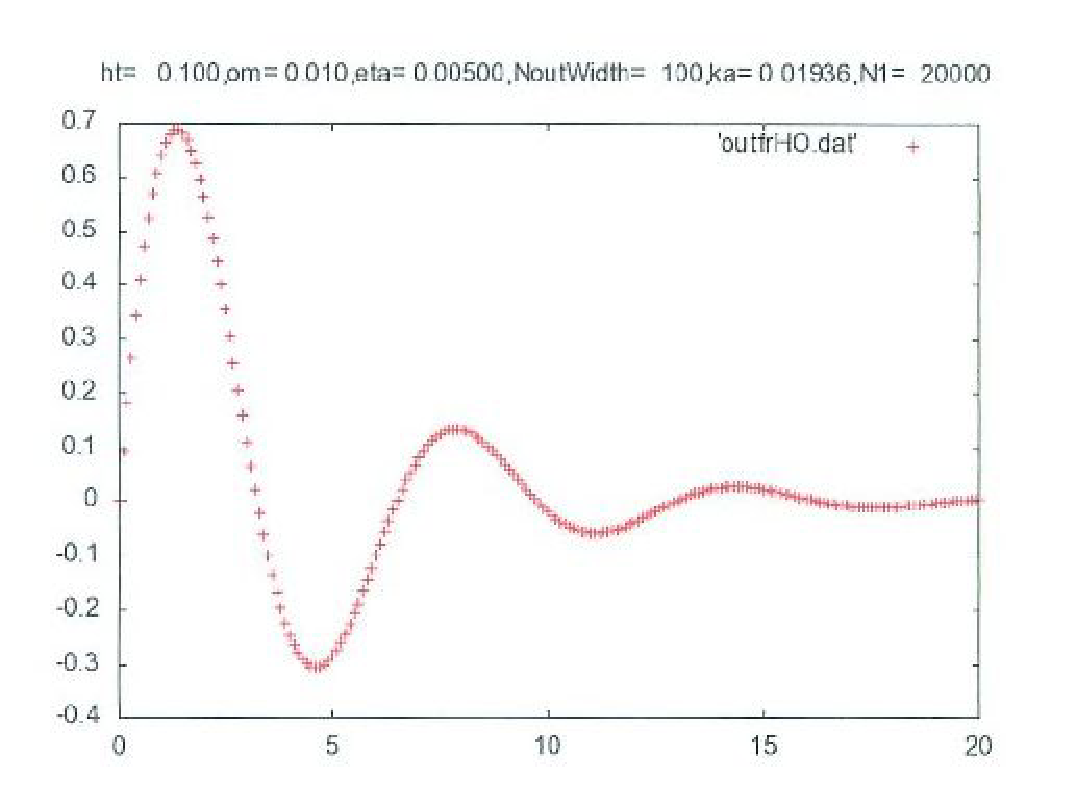}
\end{center}
\label{MovfrHO}
\end{figure}

\begin{figure}
\caption{
The velocity $v_n=(x_n - x_{n-1})/h$ of the frictional harmonic oscillator Fig.\ref{HOmodel} with the parameters (\ref{qm10}) (Elasticity dominate).  
        }
\begin{center}
\includegraphics[height=6cm]{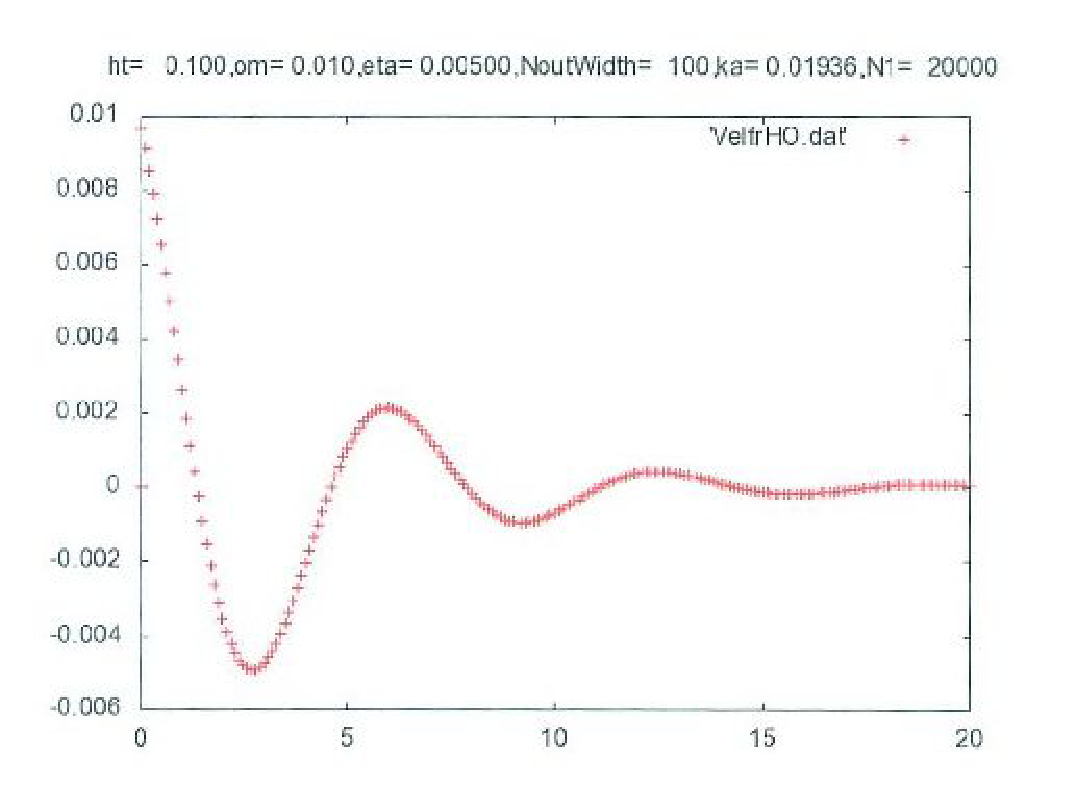}
\end{center}
\label{VelfrHO}
\end{figure}

\begin{figure}
\caption{
The system energy (\ul{SysE}$_n$, Fig.\ref{MarEfrHO}) plus the dissipative energy 
 (\ul{DisE}$_n$, Fig.\ref{EnefrHO})  of the frictional harmonic oscillator Fig.\ref{HOmodel}
 with the parameters (\ref{qm10}) (Elasticity dominate).   }
\begin{center}
\includegraphics[height=6cm]{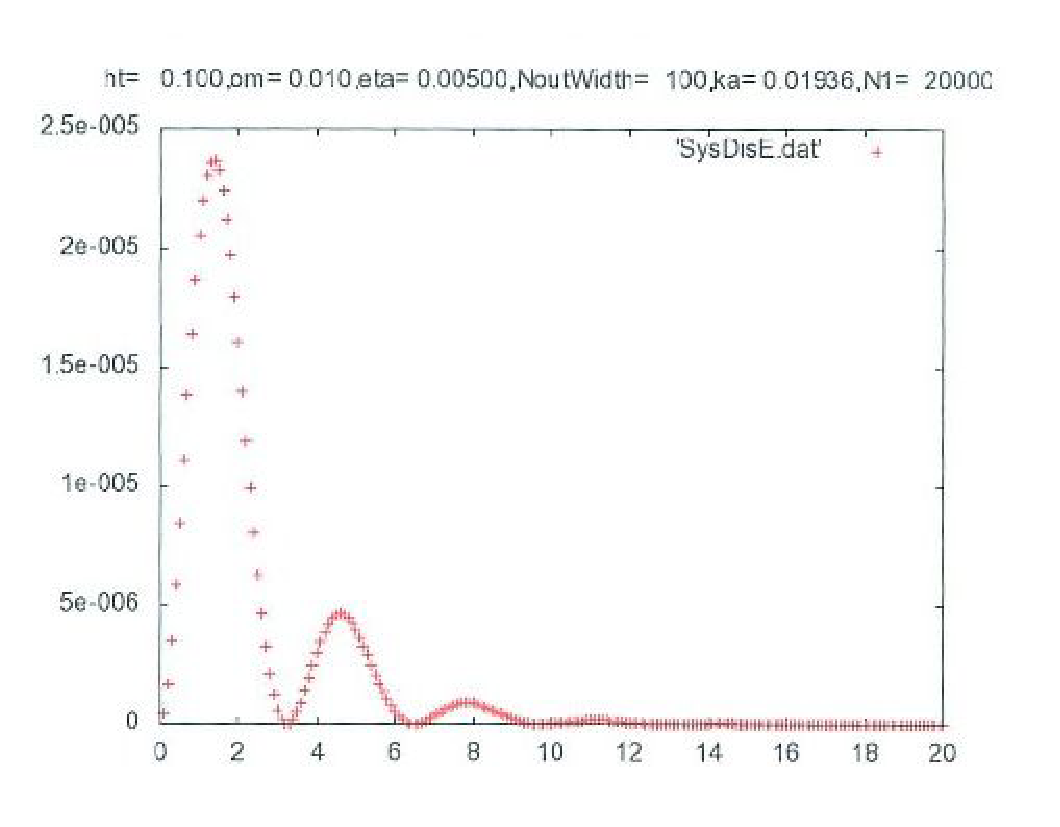}
\end{center}
\label{SysDisE}
\end{figure}

The step-wise solution (\ref{qm2}), $x_n$, is shown in Fig.\ref{MovfrHO}.  
It reproduces the analytic solution.
\bea
x(t)= \e^{-\eta' t/2}\sin(\sqrt{4\om^2-{\eta'}^2}~t/2)\com\nn 
0\leq t\leq 2000\com\q 
x(0)=0\com\q \xdot(0)=\sqrt{4\om^2-{\eta'}^2}/2=(1.94\times 10^{-2})/2
\pr
\label{Sim1}
\eea 
The velocity $v_n=(x_n - x_{n-1})/h$ is shown in Fig.\ref{VelfrHO}. 
The sum of the system energy \ul{SysE}$_n$ (Fig.\ref{MarEfrHO}) and 
the dissipative energy 
 \ul{DisE}$_n$ (Fig.\ref{EnefrHO}) is shown in Fig.\ref{SysDisE}. 

\begin{figure}
\caption{
The spring-block model, (\ref{SB2b}). 
        }
\begin{center}
\includegraphics[height=6cm]{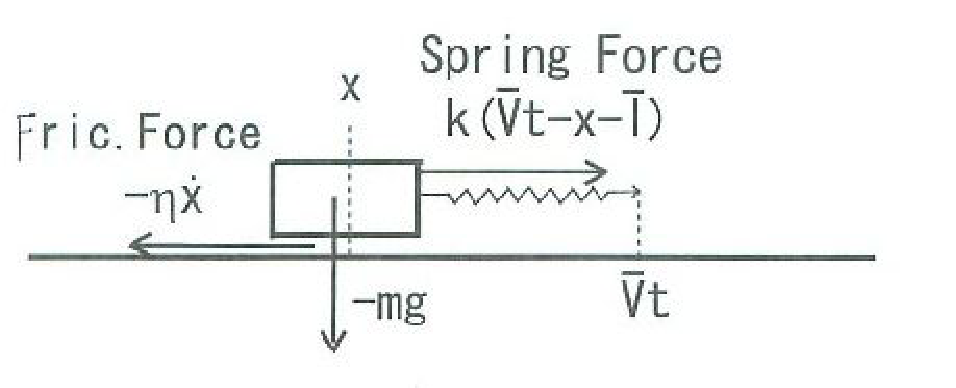}
\end{center}
\label{SBmodel}
\end{figure}
\section{Appendix D\ \ Spring-Block Model  \label{SBl} }

 In Sec.\ref{qm}, the movement of the harmonic oscillator with friction was examined. 
Here we treat the movement of a block which is pulled by the spring which moves 
at the constant speed $\Vbar$. See Fig.\ref{SBmodel}. The block 
moves on the surface with friction. We take the following $n$-th energy function 
to define the step flow. 
\bea
K_n(x)=  V(x)-hnk\Vbar x +\frac{m}{2h^2}(x-2x_{n-1}+x_{n-2})^2+
\frac{\eta}{2h}(x-x_{n-1})^2+K_n^0\com\nn
V(x)=\frac{kx^2}{2}+k\lbar x
\com
\label{SB1}
\eea 
where  $\eta$ is the friction coefficient and $m$ is the block mass. 
The potential $V(x)$ has two terms: one is the harmonic oscillator
with the spring constant $k$, and the other is the linear term of x with 
a new parameter $\lbar$ (the natural length of the spring). 
$\Vbar$ is the velocity (constant) with which the front-end of the spring moves. 
$K_n^0$ is a constant which does not depend on $x$.  It will be fixed later. 
The $n$-th step $\xn$ is determined by the energy minimum principle: 
$\del K_n(x)|_{x=\xn}=0$. 
\bea
\frac{k}{m}(\xn+\lbar-nh\Vbar)
+\frac{\eta}{m}\frac{1}{h}(\xn-x_{n-1})+\frac{1}{h^2}(\xn-2x_{n-1}+x_{n-2})=0\q\mbox{or}\nn
\xn=\frac{\om^2(-\lbar+nh\Vbar)+\frac{\etap}{h}x_{n-1}+\frac{1}{h^2}(2x_{n-1}-x_{n-2})}
{\om^2+\frac{\eta'}{h}+\frac{1}{h^2}},\ 
\om\equiv \sqrt{\frac{k}{m}}\ ,\ \etap\equiv \frac{\eta}{m}
\pr
\label{SB2}
\eea 
For the continuous limit: $h\ra 0, nh=t_n\ra t, (\xn-x_{n-1})/h\ra\xdot, 
(\xn-2x_{n-1}+x_{n-2})/h^2\ra\xddot$, the above recursion relation reduces to 
the following differential equation. 
\footnote{
This equation is called spring-block model and is used to explain 
some aspects (stick-slip motion, etc) of the earthquake\cite{SItribint15}. 
}
\bea
m\xddot=k(\Vbar t-x-\lbar)-\eta\xdot
\pr
\label{SB2b}
\eea 

  We keep the step-wise approach. 
The system energy given by $K_n(\xn)$. Taking the constant term $K_n^0$ as 
\bea
K_n^0=-V(\xn)-\frac{m}{2h^2}(\xn-2x_{n-1}+x_{n-2})^2+hnk\Vbar \xn+V(x_0)+\frac{m}{2h^2}(x_1-x_0)^2
,
\label{SB3}
\eea 
the energy is given as
\bea
K_n(\xn)=  
\frac{\eta}{2h}(\xn-x_{n-1})^2+V(x_0)+\frac{m}{2h^2}(x_1-x_0)^2  
\com
\label{SB4}
\eea 
This is the same as (\ref{qm9}). 
We have taken the constant term $K^0_n$, (\ref{SB3}), in such a way that the system keeps 
the constant energy when the energy dissipation does not occur 
.  The first three terms in (\ref{SB3}) comes from the following relation derived from (\ref{SB2b}). 
\bea
\left[ \frac{m}{2}\xdot^2+\frac{k}{2} x^2+k \lbar x \right]_0^t  +\left[ -k\Vbar x t \right]_0^t
+\int_0^t(\eta\xdot^2+k\Vbar x) dt =0
\pr
\label{SB5}
\eea 
We use the equation :\ $
\int_0^t(-k\Vbar \xdot t) dt = 
\left[ -k\Vbar x t \right]_0^t + \int_0^t(k\Vbar x) dt 
                                $. 
The terms bracketed above correspond to the first three terms in $K_n(x_n)$ which is 
obtained by taking $x=x_n$ in (\ref{SB1}). 
Those terms are regarded as Markovian and canceled by the first three terms in (\ref{SB3}).  

\begin{figure}
\caption{
Spring-Block Model, Movement (\ref{SB2}),  
$h$=0.0001,$\om=\sqrt{k/m}$=10.0, $\eta'=\eta/m$=1.0, $\Vbar$=1.0, $\lbar$=1.0, total step no =20000. 
The horizontal axis is $n\times h\times \om$. 
The step-wise solution (\ref{SB2}) correctly reproduces the analytic solution: 
$x(t)=\e^{-\eta' t/2}\Vbar\{ ({\eta'}^2/2\om^2-1)(\sin\Om t)/\Om +(\eta'/\om^2)\cos\Om t \}
-\lbar+\Vbar (t-\eta'/\om^2)\com \Om=(1/2)\sqrt{4\om^2-{\eta'}^2}=9.99\com
\ 0\leq t\leq 2\com\q x(0)=-\lbar,\ \xdot(0)=0$. 
        }
\begin{center}
\includegraphics[height=12cm]{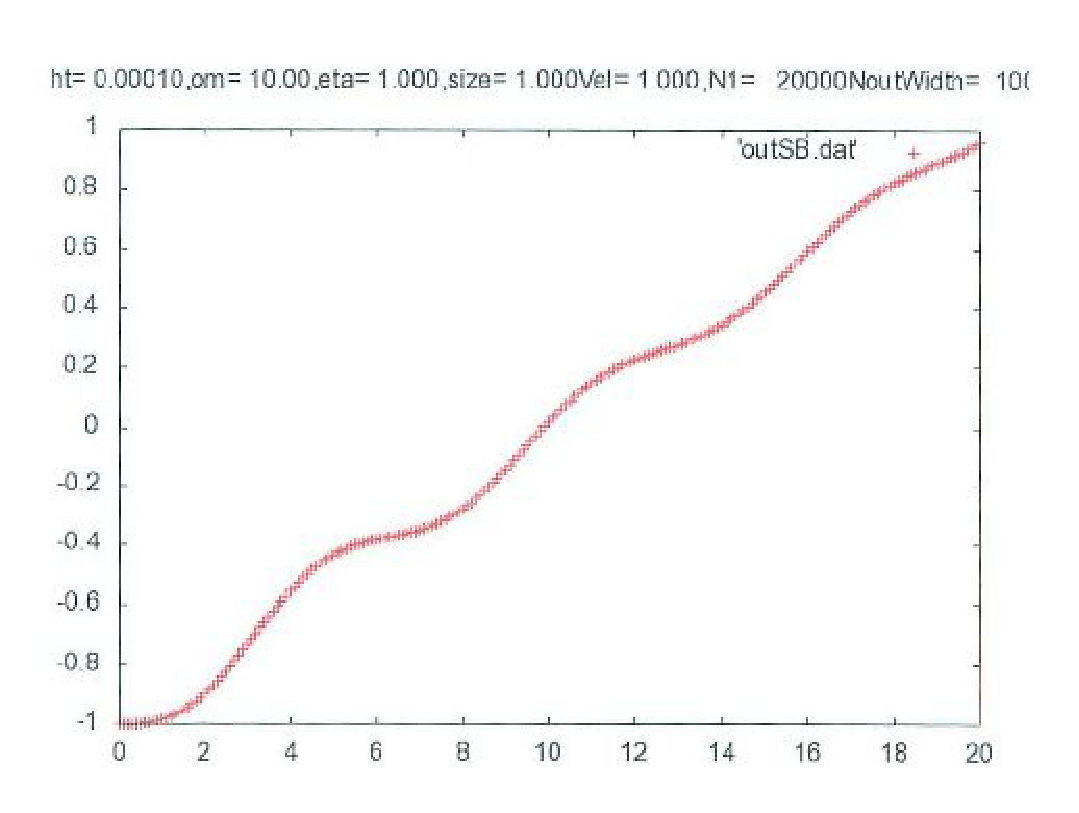}    
\end{center}
\label{MovSB}
\end{figure}

\begin{figure}
\caption{
Velocity $(x_n-x_{n-1})/h\equiv v_n$,  
$h$=0.0001,$\om=\sqrt{k/m}$=10.0, $\eta'=\eta/m$=1.0, $\Vbar$=1.0, $\lbar$=1.0, total step no =20000. 
        }
\begin{center}
\includegraphics[height=12cm]{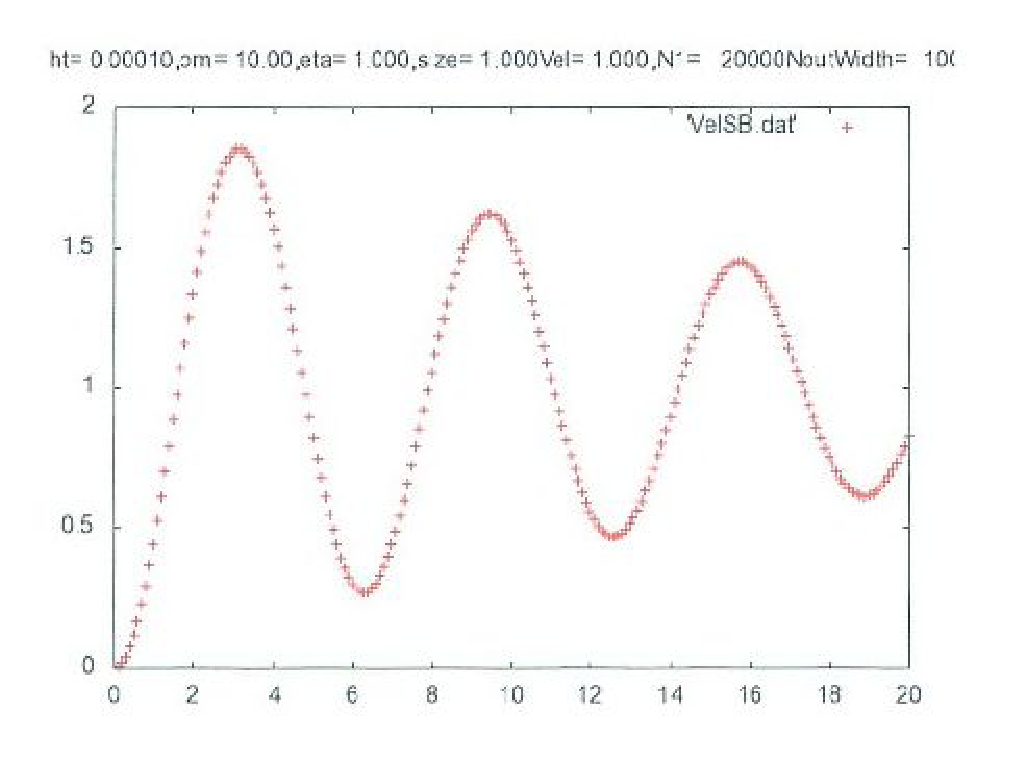} 
\end{center}
\label{VelSB}
\end{figure}

\begin{figure}
\caption{
Spring-Block Model, Energy Change (\ref{SB4}), 
$h$=0.0001,$\om=\sqrt{k/m}$=10.0, $\eta'=\eta/m$=1.0, $\Vbar$=1.0, $\lbar$=1.0, total step no =20000.
        }
\begin{center}
\includegraphics[height=12cm]{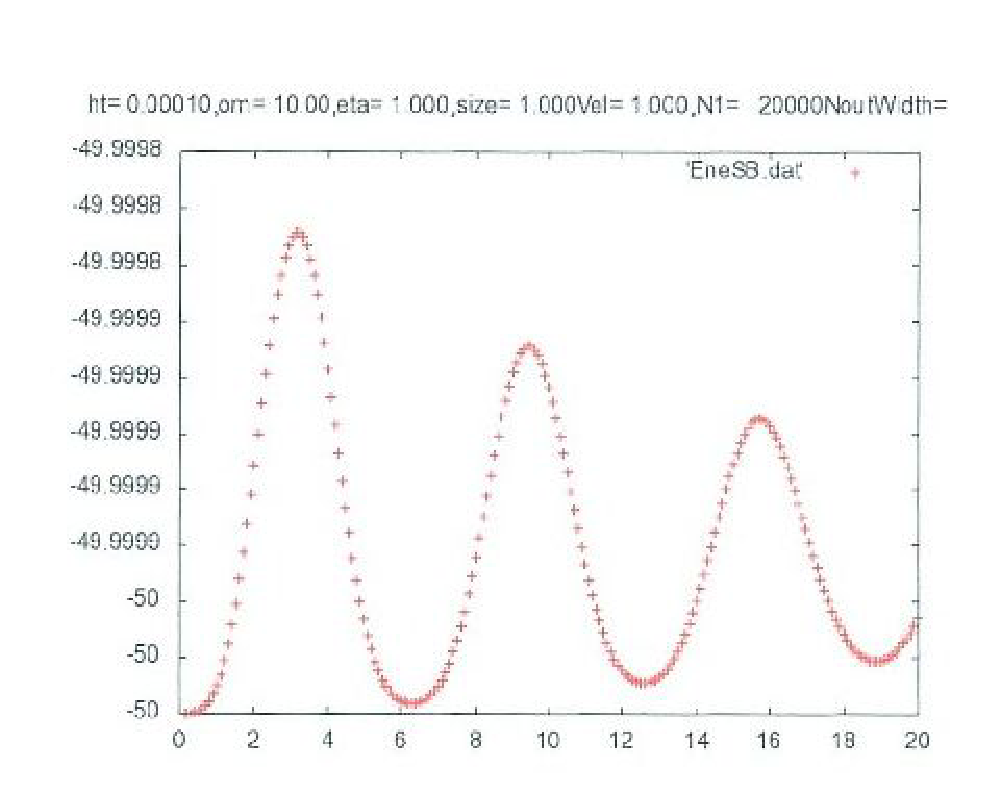}   
\end{center}
\label{EneSB}
\end{figure}

The graphs of movement ($\xn$, eq.(\ref{SB2})) and energy change ($K_n(\xn)$, eq.(\ref{SB4})) are shown in Fig.\ref{MovSB} and Fig.\ref{EneSB} respectively. 
The velocity change ($v_n\equiv (x_n-x_{n-1})/h$) is also shown in Fig.\ref{VelSB}. 
From the graphs, 
we see this system does 
the {\it stick-slip} motion. The stick regions correspond to the neighbor of the 
local minimums in the velocity change Fig.\ref{VelSB}.   
The system oscillates periodically in the velocity (Fig.\ref{VelSB}) and in the energy (Fig.\ref{EneSB}). The oscillation amplitudes decay as the step goes (relaxation). 
Finally the system 
reaches the steady energy-state as $n\ra \infty$, (\ref{coord6b}).

\section{Acknowledgment\label{ack} }
\ This work has finished during the author's stay in DAMTP, Cambridge. 
He thanks all members for the hospitality, especially G. W. Gibbons for 
comments about this work. He also thanks N. Kikuchi (Keio Univ., Japan) 
for introducing his theory\cite{Kikuchi}. The present research starts from it. Finally 
the author thanks K. Mori (Saitama Univ., Japan) for the continual advices about 
mathematics and physics which leads to this work. 
This research project is financially supported by University of Shizuoka. 
(
March 2013
)

The content is now improved and was reported in some workshops and conferences\cite{APPC1308}.

\end{document}

%% file: revmacro.tex
\DeclareMathAlphabet{\mib}{OML}{cmm} {b}{it}
\definecolor{cyan}{cmyk}{1,0,0,0}
\definecolor{lightcyan}{cmyk}{0.5,0,0,0}
\definecolor{pastelcyan}{cmyk}{0.25,0,0,0}
\definecolor{magenta}{cmyk}{0,1,0,0}
\definecolor{yellow}{cmyk}{0,0,1,0}
\definecolor{lightyellow}{cmyk}{0,0,0.5,0}
\definecolor{pastelyellow}{cmyk}{0,0,0.25,0}
\definecolor{black}{cmyk}{0,0,0,1}
\definecolor{darkgray}{cmyk}{0,0,0,0.75}
\definecolor{gray}{cmyk}{0,0,0,0.5}
\definecolor{lightgray}{cmyk}{0,0,0,0.25}
\definecolor{white}{cmyk}{0,0,0,0}
\definecolor{red}{cmyk}{0,1,1,0}
\definecolor{orange}{cmyk}{0,0.5,1,0}
\definecolor{scarlet}{cmyk}{0,1,0.5,0}
\definecolor{brown}{cmyk}{0.5,0.75,1,0}
\definecolor{camel}{cmyk}{0.25,0.375,0.5,0}
\definecolor{cream}{cmyk}{0,0.2,0.3,0}
\definecolor{green}{cmyk}{1,0,1,0}
\definecolor{lightgreen}{cmyk}{0.5,0,0.5,0}
\definecolor{pastelgreen}{cmyk}{0.25,0,0.25,0}
\definecolor{mossgreen}{cmyk}{0.64,0.4,1,0}
\definecolor{yellowgreen}{cmyk}{0.5,0,1,0}
\definecolor{skyblue}{cmyk}{0.4,0.16,0,0}
\definecolor{royal}{cmyk}{1.0,0.5,0,0}
\definecolor{navyblue}{cmyk}{0.9,0.75,0.5,0}
\definecolor{blue}{cmyk}{1,1,0,0}
\definecolor{lightblue}{cmyk}{0.5,0.5,0,0}
\definecolor{lavender}{cmyk}{0.25,0.25,0,0}
\definecolor{violet}{cmyk}{0.75,1,0.25,0}
\definecolor{purple}{cmyk}{0.5,1,0.5,0}
\definecolor{pink}{cmyk}{0,0.5,0,0}
\definecolor{pastelpink}{cmyk}{0,0.25,0,0}
\def\e{{\rm e}}
\def\gtsim{\mathrel{\hbox{\raise0.2ex
\hbox{$>$}\kern-0.75em\raise-0.9ex\hbox{$\sim$}}}}
\def\ltsim{\mathrel{\hbox{\raise0.2ex
\hbox{$<$}\kern-0.75em\raise-0.9ex\hbox{$\sim$}}}}
\def\llt{\mathrel{<\kern-0.5em <}}   
\def\ggt{\mathrel{>\kern-0.5em >}}   

\def\Tr{{\rm Tr}}
\def\half{{1\over2}}
\def\kslash{k\kern-0.55em\raise 0.14ex\hbox{/}}
\def\Aslash{A\kern-0.6em\raise 0.14ex\hbox{/}}
\def\dslash{\del\kern-0.55em\raise 0.14ex\hbox{/}}
\def\therefore{ \hskip 0.0pt \raise0.1ex\hbox{.} \mkern -0.24mu \raise 1.1ex\hbox{.} \mkern -0.25mu \raise 0.1ex\hbox{.} \   } 
\def\because{ \hskip 0.0pt \raise 1.1ex\hbox{.} \mkern -0.24mu \raise 0.1ex\hbox{.} \mkern -0.25mu \raise 1.1ex\hbox{.} \  } 